\documentclass[twocolumn]{aastex631}
\let\tablenum\relax
\usepackage{siunitx}
\usepackage{tabularx}
\usepackage{amsmath}
\usepackage{multirow}

\begin{document}

\title{ALMA Lensing Cluster Survey: Physical characterization of near-infrared-dark intrinsically faint ALMA sources at z=2-4}

\author[0000-0002-0498-5041]{Akiyoshi Tsujita}
\affiliation{Institute of Astronomy, Graduate School of Science, The University of Tokyo, 2-21-1 Osawa, Mitaka, Tokyo 181-0015, Japan}

\author[0000-0002-4052-2394]{Kotaro Kohno}
\affiliation{Institute of Astronomy, Graduate School of Science, The University of Tokyo, 2-21-1 Osawa, Mitaka, Tokyo 181-0015, Japan}
\affiliation{Research Center for the Early Universe, Graduate School of Science, The University of Tokyo, 7-3-1 Hongo, Bunkyo-ku, Tokyo 113-0033, Japan}

\author[0009-0006-1731-6927]{Shuo Huang}
\affiliation{National Astronomical Observatory of Japan, 2-21-1 Osawa, Mitaka, Tokyo 181-8588, Japan}
\affiliation{Department of Physics, Graduate School of Science, Nagoya University, Furocho, Chikusa, Nagoya 464-8602, Japan}

\author[0000-0003-3484-399X]{Masamune Oguri}
\affiliation{Center for Frontier Science, Chiba University, 1-33 Yayoi-cho, Inage-ku, Chiba 263-8522, Japan}
\affiliation{Department of Physics, Graduate School of Science, Chiba University, 1-33 Yayoi-Cho, Inage-Ku, Chiba 263-8522, Japan}

\author[0000-0001-9728-8909]{Ken-ichi Tadaki}
\affiliation{Graduate School of Engineering, Hokkai-Gakuen University, 1-1, Nishi 11-chome, Minami 26-jo, Chuo-ku, Sapporo-shi, Hokkaido 064-0926, Japan}

\author[0000-0003-3037-257X]{Ian Smail}
\affiliation{Centre for Extragalactic Astronomy, Department of Physics, Durham University, South Road, Durham DH1 3LE, UK}

\author[0000-0003-1937-0573]{Hideki Umehata}
\affiliation{Institute for Advanced Research, Nagoya University, Furocho, Chikusa, Nagoya 464-8602, Japan}
\affiliation{Department of Physics, Graduate School of Science, Nagoya University, Furocho, Chikusa, Nagoya 464-8602, Japan}
\affiliation{Cahill Center for Astronomy and Astrophysics, California Institute of Technology, 1200 E California Blvd, MC 249-17, Pasadena, CA 91125, USA}

\author[0000-0003-1262-7719]{Zhen-Kai Gao}
\affiliation{Institute of Astronomy and Astrophysics, Academia Sinica, Taipei 10617, Taiwan}
\affiliation{Graduate Institute of Astronomy, National Central University, Zhongli, Taoyuan 32001, Taiwan}

\author[0000-0003-2588-1265]{Wei-Hao Wang}
\affiliation{Institute of Astronomy and Astrophysics, Academia Sinica, Taipei 10617, Taiwan}

\author[0000-0002-4622-6617]{Fengwu Sun}
\affiliation{Center for Astrophysics $|$ Harvard \& Smithsonian, 60 Garden St., Cambridge MA 02138 USA}
\affiliation{Steward Observatory, University of Arizona, 933 N. Cherry Avenue, Tucson, AZ 85721, USA}

\author[0000-0001-7201-5066]{Seiji Fujimoto}
\affiliation{Department of Astronomy, The University of Texas at Austin, Austin, TX, USA}
\affiliation{Cosmic Dawn Center (DAWN), Jagtvej 128, DK2200 Copenhagen N, Denmark}
\affiliation{Niels Bohr Institute, University of Copenhagen, Jagtvej 128, DK2200 Copenhagen N, Denmark}

\author[0000-0002-2504-2421]{Tao Wang}
\affiliation{Key Laboratory of Modern Astronomy and Astrophysics (Nanjing University), Ministry of Education, Nanjing 210093, PR China}

\author[0000-0001-6653-779X]{Ryosuke Uematsu}
\affiliation{Department of Astronomy, Kyoto University, Kyoto 606-8502, Japan}

\author[0000-0002-8726-7685]{Daniel Espada}
\affiliation{Departamento de F\'{i}sica Te\'{o}rica y del Cosmos, Campus de Fuentenueva, Edificio Mecenas, Universidad de Granada, E-18071, Granada, Spain}
\affiliation{Instituto Carlos I de F\'{i}sica Te\'{o}rica y Computacional, Facultad de Ciencias, E-18071, Granada, Spain}

\author[0000-0001-6477-4011]{Francesco Valentino}
\affiliation{European Southern Observatory, Karl-Schwarzschild-Str. 2, D-85748, Garching, Germany}
\affiliation{Cosmic Dawn Center (DAWN), Denmark}

\author[0000-0003-3139-2724]{Yiping Ao}
\affiliation{Purple Mountain Observatory and Key Laboratory for Radio Astronomy, Chinese Academy of Sciences, Nanjing, PR China}
\affiliation{School of Astronomy and Space Science, University of Science and Technology of China, Hefei, PR China}

\author[0000-0002-8686-8737]{Franz E. Bauer}
\affiliation{Instituto de Astrof{\'{\i}}sica, Facultad de F{\'{i}}sica, Pontificia Universidad Cat{\'{o}}lica de Chile, Campus San Joaquín, Av. Vicuña Mackenna 4860, Macul Santiago, Chile, 7820436} 
\affiliation{Centro de Astroingenier{\'{\i}}a, Facultad de F{\'{i}}sica, Pontificia Universidad Cat{\'{o}}lica de Chile, Campus San Joaquín, Av. Vicuña Mackenna 4860, Macul Santiago, Chile, 7820436} 
\affiliation{Millennium Institute of Astrophysics, Nuncio Monse{\~{n}}or S{\'{o}}tero Sanz 100, Of 104, Providencia, Santiago, Chile}

\author[0000-0001-6469-8725]{Bunyo Hatsukade}
\affiliation{National Astronomical Observatory of Japan, 2-21-1 Osawa, Mitaka, Tokyo 181-8588, Japan}
\affiliation{Graduate Institute for Advanced Studies, SOKENDAI, Osawa, Mitaka, Tokyo 181-8588, Japan}
\affiliation{Institute of Astronomy, Graduate School of Science, The University of Tokyo, 2-21-1 Osawa, Mitaka, Tokyo 181-0015, Japan}

\author[0000-0002-1639-1515]{Fumi Egusa}
\affiliation{Institute of Astronomy, Graduate School of Science, The University of Tokyo, 2-21-1 Osawa, Mitaka, Tokyo 181-0015, Japan}

\author[0000-0003-0563-067X]{Yuri Nishimura}
\affiliation{Institute of Astronomy, Graduate School of Science, The University of Tokyo, 2-21-1 Osawa, Mitaka, Tokyo 181-0015, Japan}

\author[0000-0002-6610-2048]{Anton M. Koekemoer}
\affiliation{Space Telescope Science Institute, 3700 San Martin Dr., Baltimore, MD 21218, USA} 

\author[0000-0001-7144-7182]{Daniel Schaerer}
\affiliation{
Observatoire de Gen\'eve, Universit\'e de Gen\'eve, 51 Ch. des Maillettes, 1290 Versoix, Switzerland
}
\affiliation{CNRS, IRAP, 14 Avenue E. Belin, 31400 Toulouse, France
}

\author[0000-0003-3021-8564]{Claudia Lagos}
\affiliation{International Centre for Radio Astronomy Research (ICRAR), M468, University of Western Australia, 35 Stirling Hwy, Crawley, \\WA 6009, Australia}
\affiliation{ARC Centre of Excellence for All Sky Astrophysics in 3 Dimensions (ASTRO 3D)}
\affiliation{Cosmic Dawn Center (DAWN), Denmark}

\author[0000-0003-0348-2917]{Miroslava Dessauges-Zavadsky}
\affiliation{Department of Astronomy, University of Geneva, Chemin Pegasi 51, 1290 Versoix, Switzerland}

\author[0000-0003-2680-005X]{Gabriel Brammer}
\affiliation{
Cosmic Dawn Center (DAWN), Denmark
}
\affiliation{
Niels Bohr Institute, University of Copenhagen, Jagtvej 128, DK2200 Copenhagen N, Denmark
}

\author[0000-0001-8183-1460]{Karina Caputi}
\affiliation{
Kapteyn Astronomical Institute, University of Groningen, P.O. Box 800, 9700AV Groningen, The Netherlands
}
\affiliation{
Cosmic Dawn Center (DAWN), Denmark
}

\author[0000-0003-1344-9475]{Eiichi Egami}
\affiliation{
Steward Observatory, University of Arizona, 933 N. Cherry Ave, Tucson, AZ 85721, USA
}

\author[0000-0003-3926-1411]{Jorge Gonz\'alez-L\'opez}
\affiliation{
N\'ucleo de Astronom\'ia de la Facultad de Ingenier\'ia y Ciencias, Universidad Diego Portales, Av. Ejército Libertador 441, Santiago, Chile
}
\affiliation{
Las Campanas Observatory, Carnegie Institution of Washington, Casilla 601, La Serena, Chile
}

\author[0000-0002-3405-5646]{Jean-Baptiste Jolly}
\affiliation{Department of Space, Earth and Environment, Chalmers University of Technology, Onsala Space Observatory, SE-439 92 Onsala, Sweden
} 
\affiliation{Max-Planck-Institut f\"{u}r extraterrestrische Physik, 85748 Garching, Germany;
} 

\author[0000-0002-7821-8873]{Kirsten K. Knudsen}
\affiliation{Department of Space, Earth and Environment, Chalmers University of Technology, Onsala Space Observatory, SE-43992 Onsala, Sweden}

\author[0000-0002-5588-9156]{Vasily Kokorev}
\affiliation{
Kapteyn Astronomical Institute, University of Groningen, P.O. Box 800, 9700AV Groningen, The Netherlands
}
\affiliation{
Cosmic Dawn Center (DAWN), Denmark
}
\affiliation{
Niels Bohr Institute, University of Copenhagen, Jagtvej 128, DK2200 Copenhagen N, Denmark
}

\author[0000-0002-4872-2294]{Georgios E. Magdis}
\affiliation{
Cosmic Dawn Center (DAWN), Denmark
}
\affiliation{DTU-Space, Technical University of Denmark, Elektrovej 327, 2800, Kgs. Lyngby, Denmark}
\affiliation{Niels Bohr Institute, University of Copenhagen, Jagtvej 128, 2200, Copenhagen N, Denmark}

\author[0000-0002-1049-6658]{Masami Ouchi}
\affiliation{National Astronomical Observatory of Japan, 2-21-1 Osawa, Mitaka, Tokyo 181-8588, Japan}
\affiliation{Institute for Cosmic Ray Research, The University of Tokyo, 5-1-5 Kashiwanoha, Kashiwa, Chiba 277-8582, Japan}
\affiliation{Kavli Institute for the Physics and Mathematics of the Universe (WPI), University of Tokyo, Kashiwa, Chiba 277-8583, Japan}

\author[0000-0003-3631-7176]{Sune Toft}
\affiliation{Cosmic Dawn Center (DAWN), Denmark}
\affiliation{Niels Bohr Institute, University of Copenhagen, Jagtvej 128, DK2200 Copenhagen N, Denmark}

\author[0000-0002-5077-881X]{John F. Wu}
\affiliation{Space Telescope Science Institute, 3700 San Martin Dr., Baltimore, MD 21218, USA}
\affiliation{Center for Astrophysical Sciences, Johns Hopkins University, 3400 N Charles St., Baltimore, MD 21218, USA}

\author[0000-0002-0350-4488]{Adi Zitrin}
\affiliation{
Physics Department, Ben-Gurion University of the Negev, P.O. Box 653, Be’er-sheva 8410501, Israel
}







\begin{abstract} 
We present results from Atacama Large Millimeter/submillimeter Array (ALMA) spectral line-scan observations at 3-mm and 2-mm bands of three near-infrared-dark (NIR-dark) galaxies behind two massive lensing clusters MACS J0417.5-1154 and RXC J0032.1+1808. Each of these three sources is a (sub)mm faint (de-lensed $S_{\rm 1.2 mm}$ $<$ 1 mJy) triply lensed system originally discovered in the ALMA Lensing Cluster Survey. We have successfully detected CO and [C~I] emission lines and confirmed that their spectroscopic redshifts are $z=3.652$, 2.391, and 2.985. By utilizing a rich multi-wavelength data set, we find that the NIR-dark galaxies are located on the star formation main sequence in the intrinsic stellar mass range of log ($M_*$/$M_\odot$) = 9.8 – 10.4, which is about one order of magnitude lower than that of typical submillimeter galaxies (SMGs). These NIR-dark galaxies show a variety in gas depletion times and spatial extent of dust emission. 
One of the three is a normal star-forming galaxy with gas depletion time consistent with a scaling relation, and its infrared surface brightness is an order of magnitude smaller than that of typical SMGs.
Since this galaxy has an elongated axis ratio of $\sim 0.17$, we argue that normal star-forming galaxies in an edge-on configuration can be heavily dust-obscured. This implies that existing deep WFC3/F160W surveys may miss a fraction of typical star-forming main-sequence galaxies due to their edge-on orientation.
\end{abstract}



\section{Introduction} \label{sec:intro}
Dust affects the spectral energy distribution (SED) of galaxies by absorbing and scattering ultraviolet (UV) and optical emission, mainly produced by young stars, and re-emitting the energy at the far-infrared (FIR) and sub-millimeter (sub-mm) wavelengths.
Consequently, dusty galaxies at $z\gtrsim 2$ typically appear very faint at observed-frame optical/near-infrared (NIR) due to dust obscuration.
Instead, their dust continuum is prominent at FIR/sub-mm band and some extreme ones are bright enough (e.g., $S_{\SI{870}{\mu m}}>\mathrm{several\  mJy}$) to be detected with single-dish telescopes. They are referred to as submillimeter galaxies (SMGs; \citealt{Hodge2020} for a review).
Since the first detection of high-redshift SMGs, many have been found to be completely obscured in the rest-frame UV/optical (or observed-frame NIR) due to strong dust attenuation, dubbed NIR-dark galaxies (e.g., \citealt{Smail1999, dacunha2015, Yamaguchi2019, Casey2019, Umehata2020, Dudzeviciute2020, Riechers2020, Sun2021, zavala2021, Smail2021, Shu2022}).
Similar terms, such as $HST$-dark and $H/K$-dropout, are also used to describe these dust-obscured sources.

The advent of the Atacama Large Millimeter/submillimeter Array (ALMA) has enabled us to uncover the faint ($S_{\mathrm{1.2mm}} < \SI{1}{mJy}$) (sub-)mm-selected galaxies, that are significantly fainter (up to $\sim100$ times) than the classical SMGs.
Hereafter, we will use the term dusty star-forming galaxies (DSFGs) to more generally refer to dusty galaxies including these faint sources.
It has been revealed that the dust-obscured galaxies exist not only among SMG-like extreme starburst galaxies but also among more common DSFGs which lie on the main sequence of the $M_*$--SFR relation \citep{Schreiber2015}, particularly at the high mass end where $M_* > 10^{10.3} M_\odot$ (e.g., \citealt{Wang2019Nat}).
These galaxies have been detected through blind ALMA surveys (e.g., \citealt{Simpson2014, Yamaguchi2016, Fujimoto2016, Franco2018, Yamaguchi2019, Dudzeviciute2020, Gruppioni2020, Xiao2023, Fujimoto2024ApJS..275...36F,Fujimoto2023b}), follow-up ALMA observations of galaxies with very red colors in NIR range (e.g., \citealt{Caputi2012, Wang2016, Wang2019Nat, Alcalde2019}), follow-up ALMA observations of strongly-lensed galaxies that are identified in (sub)mm-single-dish surveys \citep{Pope2017ApJ...838..137P, Pope2023ApJ...951L..46P, Mizener2024ApJ...970...30M, Shu2022}, and serendipitous discoveries \citep{Hatsukade2015, Williams2019, Romano2020, Fudamoto2021}. 
Recent {\it James Webb Space Telescope} ({\it JWST}) observations have also begun to uncover a broader range of NIR-dark galaxies, including those that are less massive (e.g., \citealt{Barrufet2023, Barrufet2024arXiv240408052B, Nelson2023, Perez-Gonzalez2023ApJ, Gomez-Guijarro2023, Xiao2023arXiv230902492X, Smail2023, Kokorev2023, McKinney2023ApJ...956...72M, Sun2024}).

This normal dust-obscured population was often missed by previous (sub)mm-single-dish surveys due to its modest (sub-)mm flux, lower than the confusion limit of single-dish telescopes, and was also overlooked by the Lyman break technique which relies on the rest-frame UV light.
However, \cite{Wang2019Nat} suggested that this population constitutes the bulk of massive galaxies (from 20\% at $M_*\sim 10^{10.3}\,M_\odot$ to 80\% at $M_*\sim 10^{11.0}\,M_\odot$) at $z\sim3-6$ and is the main progenitor of the most massive quiescent galaxies at $z\sim2-3$, as well as of today’s ellipticals that reside in the central regions of massive groups and clusters (see also \citealt{Valentino2020}). 
The large number of such dusty massive galaxies in the early universe is not accounted for by current semi-analytical models, posing a challenge to our understanding of massive galaxy formation.

Despite its importance, obtaining accurate redshifts for these galaxies (either optical/NIR spectroscopy or (sub-)mm line scans) is challenging, hindering efforts to characterize the population.
This difficulty is partly due to significant extinction in the rest-frame UV and optical, where key redshift indicators like the Lyman break and \SI{4000}{\AA} break lie.
Another factor is the significantly lower (sub-)mm flux of this population compared with those of classical SMGs, which makes line scan observations for redshift search much more expensive, even with ALMA.
To date, several spectroscopic redshift identifications for the NIR-dark galaxies have been reported (e.g., \citealt{Swinbank2012, Casey2019, Williams2019, Umehata2020, Zhou2020, Riechers2020, Smail2023, Kokorev2023, Sun2024, Barrufet2024arXiv240408052B}) but almost all of them are for the SMG-class bright sources except for a few cases such as \cite{Fudamoto2021} which reported the serendipitous [C II]$_{158}$ line detection of two normal dust-obscured galaxies at $z\sim7$.
Therefore, physical characterizations based on secure redshifts remain largely unexplored for the normal NIR-dark population, particularly in terms of molecular gas properties based on CO or [C I] emission lines.

Furthermore, it remains unclear what the origin of the strong dust attenuation is, or in other words, what differentiates NIR-dark galaxies from other normal main-sequence SFGs.
\cite{Smail2021} investigated the correlation between the strength of dust attenuation ($A_V$) and other physical quantities in ALMA-identified SMGs, concluding that the correlations with dust mass and stellar mass are weak, while those with the photometric redshift and the ``compactness" of the star-forming region ($\Sigma_{\mathrm{IR}}$) show stronger links. However, they suggest that other factors such as the geometry of stars/dust may also be involved, particularly for completely dust-obscured galaxies (see also \citealt{dacunha2015, Wang2018, Cochrane2023}).
Moreover, recent {\it JWST} studies claim that the viewing angle of the galaxy (i.e., an edge-on disk) may be a contributing factor \citep{Nelson2023}, while others argue that redshift and/or stellar mass are the main factors \citep{Lagos2020, Gomez-Guijarro2023, Lorenz2023}.

In this paper, we focus on normal dust-obscured galaxies originally discovered in the ALMA Lensing Cluster Survey (ALCS). 
Thanks to lens magnification, these galaxies appear as bright as classic SMGs (detectable with typical single-dish telescopes), making them suitable for follow-up observations. 
We successfully identified spectroscopic redshifts for three sources and estimated their basic physical properties via ALMA Band 3 and 4 line scan observations. 
We then compared these properties with those of other galaxy populations.
Based on the obtained results, we discuss the evolutionary scenario and the origin of the strong dust attenuation for the normal dust-obscured galaxies.

This paper is organized as follows. Section \ref{sec:data} describes the observational data and reduction. In Section \ref{sec:results}, we present the spectral redshift identification, SED fitting results, molecular gas properties, FIR morphology, and properties of the photodissociation region of our samples. 
In Section \ref{sec:discussion}, we discuss the origin of the strong dust attenuation.
A summary is presented in Section \ref{sec:summary}.
Throughout the paper, we adopt a cosmology with $\Omega_{\mathrm{m}}=0.3$, $\Omega_\Lambda=0.7$, and $H_0=\SI{70}{km.s^{-1}.Mpc^{-1}}$ and the Chabrier initial mass function (IMF, \citealt{Chabrier2003}). 
We convert values obtained by other studies from the \cite{Salpeter1955} IMF to the Chabrier IMF by dividing SFR and stellar masses by the same factor, 1.7 \citep{Speagle2014}.
All magnitudes in this paper are expressed in the AB system \citep{Oke1974}. 

\section{Data and reduction}\label{sec:data}
\subsection{ALMA Lensing Cluster Survey}
The ALMA Lensing Cluster Survey (ALCS, \citealt{Kohno2023,Fujimoto2024ApJS..275...36F}) is a Cycle-6 ALMA large program (project code: 2018.1.00035.L; PI: K. Kohno) that targets highly magnified areas within 33 massive galaxy clusters. 
The samples are selected from the best-studied clusters drawn from the Cluster Lensing And Supernova Survey with Hubble (CLASH; \citealt{Postman2012}), Hubble Frontier Fields (HFF; \citealt{Lotz2017}), and the Reionization Lensing Cluster Survey (RELICS; \citealt{Coe2019}).
The survey employs a 15-GHz-wide spectral scan in Band 6 ($\overline{\lambda_{\mathrm{obs}}}=\SI{1.15}{mm}$, $\overline{\nu_\mathrm{obs}}=\SI{260}{GHz}$). 
The total mapping area reaches $\SI{134}{arcmin^2}$ (primary beam $> 0.3$) with a typical noise level of $1\sigma\sim\SI{60}{\mu Jy.beam^{-1}}$. 
Thanks to cluster lensing, ALCS explores a unique parameter space towards faint and wide regimes, compared to existing ALMA deep surveys. 

Observations were carried out between December 2018 and December 2019 in compact array configurations of C43-1 and C43-2 (corresponding to a beam size of $\theta\sim 1\arcsec$) fine-tuned to recover strongly lensed (i.e., spatially elongated), low surface brightness sources. 
ALCS yields 180 continuum source detections in total (including multiple images) consisting of 141 blind detections with a signal-to-noise ratio (SNR) of greater than five in the natural weighting maps and 39 prior-based detections with $\mathrm{SNR=4-5}$ which have IRAC-counterparts. 
Full descriptions of the data analysis and source catalog are given in \cite{Fujimoto2024ApJS..275...36F}.

\subsection{Optical, NIR, and FIR catalogs}
The 33 clusters were observed with {\it HST}, {\it Spitzer Space Telescope}, and {\it Herschel Space Observatory}. 

\cite{Kokorev2022} builds a set of {\it HST} and {\it Spitzer}/IRAC 3.6 and \SI{4.5}{\mu m} mosaics and photometric catalogs by reprocessing archival data in the ALCS fields.
One of the difficulties in studying the galaxy clusters and other crowded regions is that the photometries taken by low-resolution instruments such as {\it Spitzer}/IRAC suffer from the effect of blending.
To alleviate the effect, they model the {\it Spitzer} photometry by convolving the {\it HST} detection image with the {\it Spitzer} PSF using the novel {\tt GOLFIR} software.

\cite{sun2022} compiles archival data from {\it Herschel}/PACS and SPIRE observations of the 33 clusters, with most of the data coming from {\it Herschel} Lensing Survey \citep{Egami2010}.
The study includes data taken at 100 and $\SI{160}{\mu m}$ with PACS, and 250, 350, $\SI{500}{\mu m}$ with SPIRE. 
Given the coarse angular resolution of {\it Herschel} (typical beam sizes are 7.4$\arcsec$ and 11.4$\arcsec$ for the 100 and $\SI{160}{\mu m}$ bands of PACS, and 18$\arcsec$, 24$\arcsec$, and 35$\arcsec$ for the 250, 350, and $\SI{500}{\mu m}$ bands of SPIRE), the authors deblended the {\it Herschel} fluxes using secure ALMA detections from the ALCS as priors.

\subsection{Target selection}
In this work, we define NIR-dark sources as those with $H-[4.5]>2.3$, which are extremely red objects with $H$-band (F160W-band) and IRAC \SI{4.5}{\mu m} channel color, following the traditional method described in \cite{Wang2016}. 
This color selection is designed to complementarily select massive star-forming galaxies that are systematically missed by the Lyman-break selection, which is specifically designed to select young and less attenuated galaxies (e.g., \citealt{Bouwens2012}).

\begin{figure}[tbh]
 \begin{center}
  \includegraphics[width=1\linewidth]{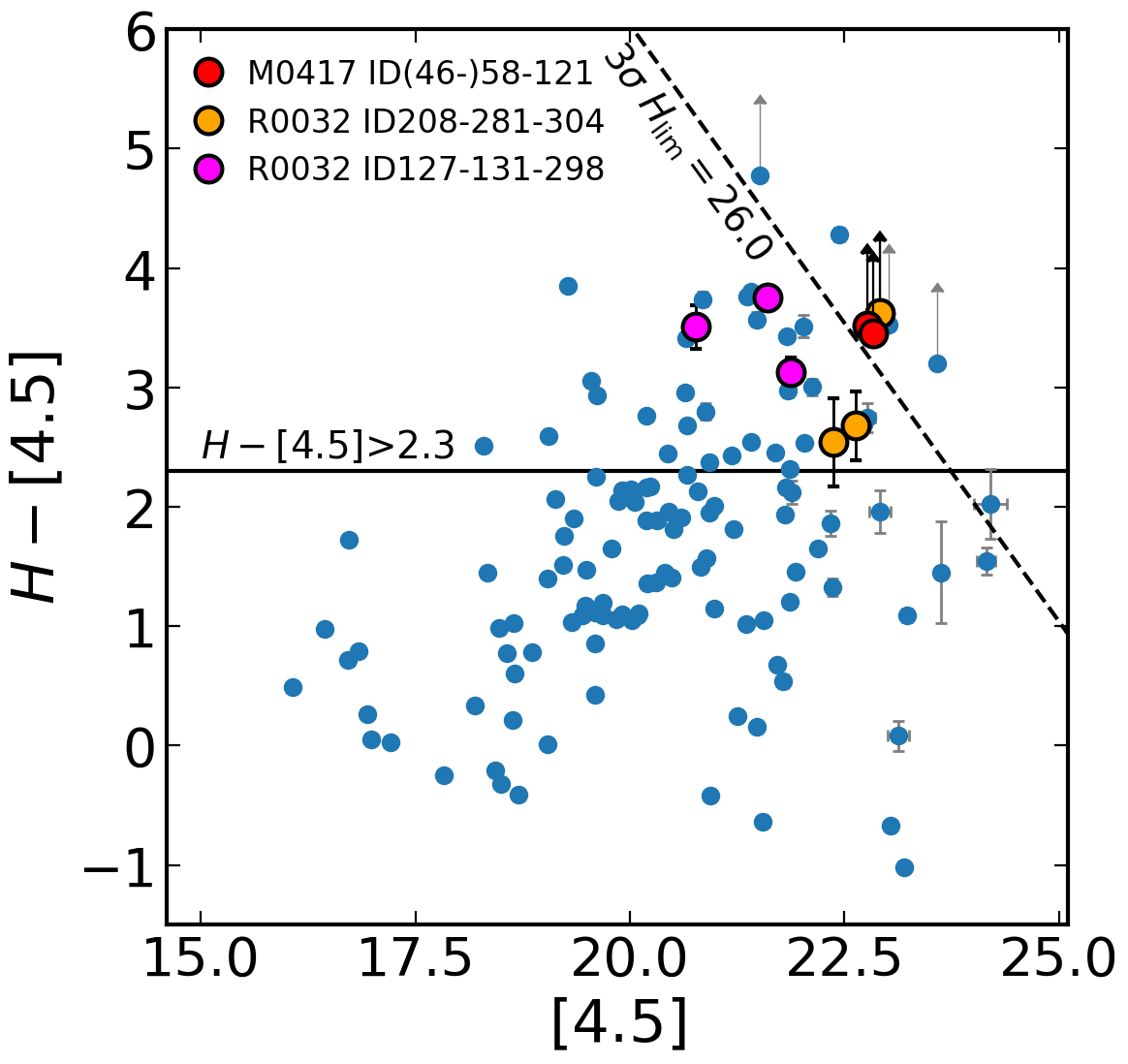}
 \end{center}
\caption{Color-magnitude diagram of ALCS sources. The black solid line represents the threshold of $H-[4.5]=2.3$. The dashed line shows the location of $H=26.0$ mag, which corresponds to the $3\sigma$ limit of the shallowest {\it HST} data in 33 ALCS cluster fields. The red, orange, and magenta dots represent three triply lensed images of our targets. However, M0417 ID46 is excluded due to contamination from a nearby bright source. For the $H$-dropout or $HST$-dark objects, the $H-[4.5]$ values are $3\sigma$ lower limits, as indicated by the arrows.}
 \label{fig:h-[4.5]}
\end{figure}

Figure \ref{fig:h-[4.5]} presents the color-magnitude diagram ($H-[4.5]$ vs. [4.5]) for the ALCS sample. 
We exclude sources that are not covered in the $H$-band, those contaminated by nearby bright objects or other artificial factors, or those identified as IRAC-dropout.
Among the NIR-dark objects, six are $H$-dropout or $HST$-dark. 
We calculated the lower limit of their $H$-band magnitude by taking three times the median uncertainty of all photometric measurements within the corresponding galaxy cluster region in the \cite{Kokorev2022} catalog. 
In the \cite{Kokorev2022} catalogs, the IRAC photometries are measured by convolving the {\it HST} detection image with the IRAC PSF. 
The IRAC photometries of the $HST$-dark sources are instead measured using a $D=3\arcsec$ aperture with some aperture corrections centered on ALMA Band 6 continuum positions \citep{Kokorev2022}. To mitigate contamination from nearby sources, we use the residual IRAC maps, wherein sources detected in the {\it HST} maps are modeled with the IRAC PSF and subtracted.

\begin{figure*}[tbh]
 \begin{center}
  \includegraphics[width=0.85\linewidth]{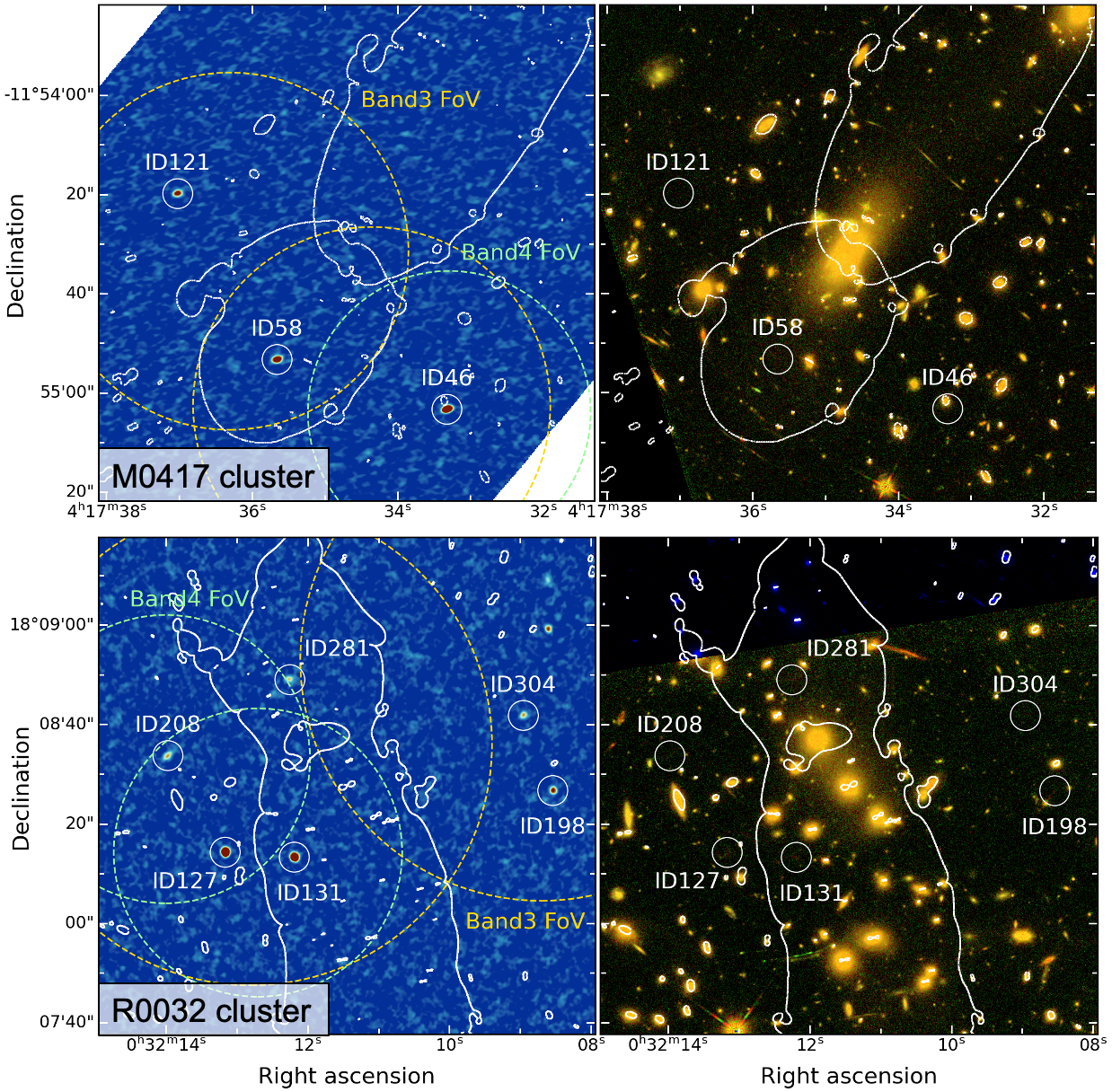}
 \end{center}
\caption{(Left) ALMA 1.2-mm continuum image of M0417 cluster. (Right) False-color HST image of M0417 cluster (red: F160W, green: F125W, blue: F814W). The top row shows the M0417 cluster and the bottom row shows the R0032 cluster. The white circles indicate the positions of multiple images of the NIR-dark galaxies. The white solid lines denote the critical lines of our fiducial lens model at $z=3.652$ and 2.985 for M0417 and R0032, respectively. The yellow and green dashed lines represent the FoV of ALMA band 3 and 4 follow-up observations described in section \ref{sec:followup}, respectively.}
 \label{fig:hst_alma}
\end{figure*}

\begin{figure*}[]
\centering
\includegraphics[width=0.9\linewidth]{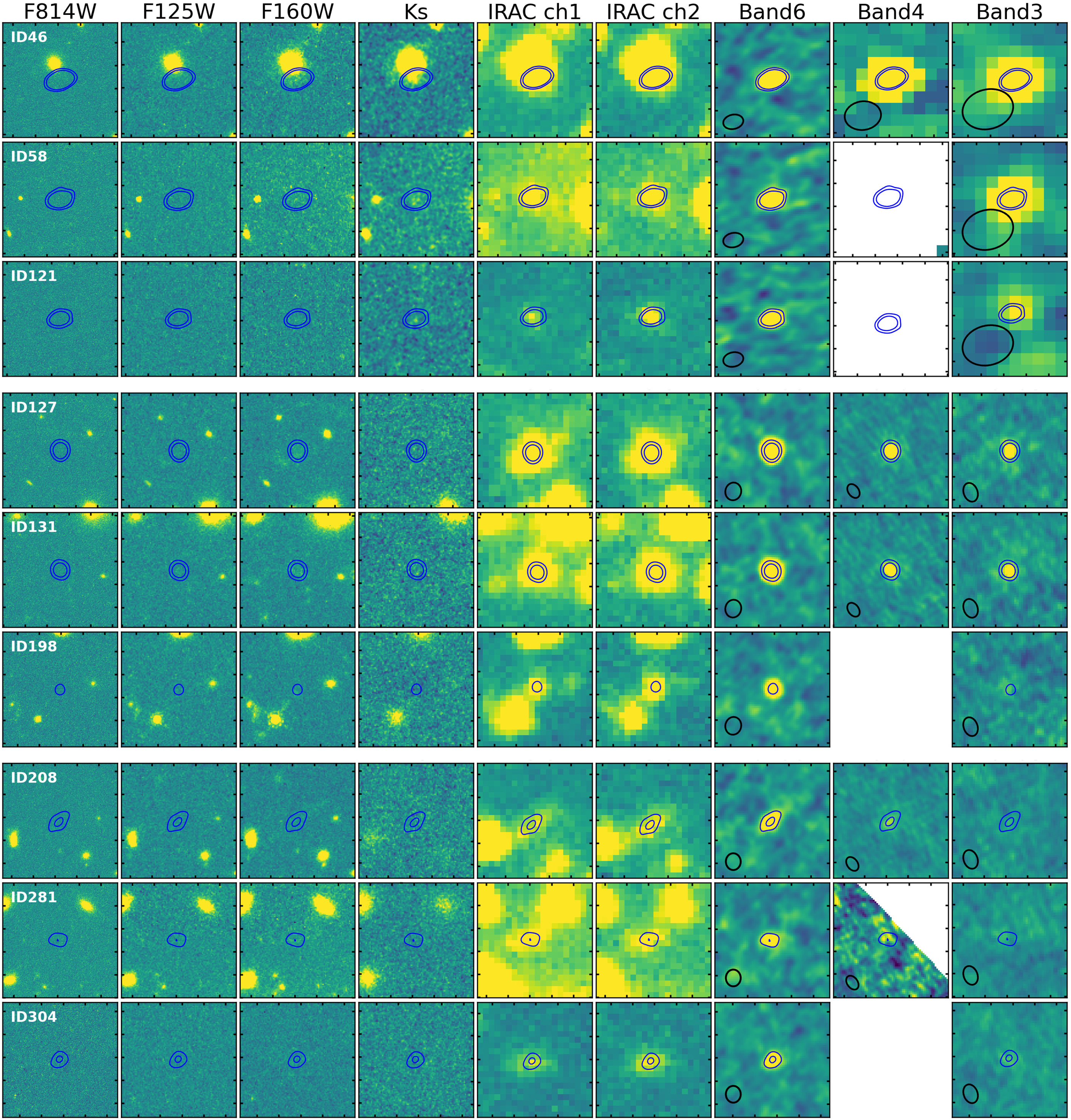}
\caption{Multi-band images for the three triply lensed systems. The blue contours denote the ALMA band 6 (1.2-mm) continuum drawn at [4, 8]$\times\sigma$ levels. The color scale is adjusted to show the $5\sigma$ range for each emission. The image size is $10\arcsec$$\times$$10\arcsec$. The black ellipses show the beam size.
\label{fig:multi_band}}
\end{figure*}

From these NIR-dark objects, we focused on three triply lensed systems: ID46, ID58, and ID121 in the MACS J0417.5-1154 (hereafter M0417) cluster region, ID127, ID131, ID198 in the RXC J0032.1+1808 (hereafter R0032) cluster region, and ID208, ID281, and ID304 in the R0032 cluster region. 
Figure \ref{fig:multi_band} shows the multi-band images for these sources.
For clarity, we will hereafter refer to the three intrinsic sources as M0417-z365, R0032-z239, and R0032-z299, and their associated triply lensed images as M0417-z365.ID46/.ID58/.ID121, R0032-z239.ID127/.ID131/.ID198, and R0032-z299.ID208/.ID281/.ID304, respectively. The ``z'' values represent the spectroscopic redshift of each source, as described in Section \ref{sec:specz}.
Note that M0417-z365.ID46 is excluded in Figure \ref{fig:h-[4.5]} due to the blending with a foreground galaxy (see Figure \ref{fig:multi_band}).
These systems were predicted to be multiple images by our fiducial lens model (section \ref{sec:lensmodel}). 
Figure \ref{fig:hst_alma} shows their ALMA 1.2 mm continuum and $HST$ images with critical lines of our fiducial lens models.
We selected them since we are able to estimate their redshifts by the lens model, which allowed us to carry out line scan observations even though the photometric redshifts are unreliable. 

\subsection{ALMA follow-up line scan observations} \label{sec:followup}
Based on the redshift estimates, we conducted follow-up line-scan observations for the three lensed systems with ALMA band 3 and 4 (project code: 2019.1.00949.S and 2021.1.01246 for M0417 and R0032 region, respectively).
Four and three different tunings were assigned to band 3 and 4 observations, respectively.
The observational parameters are summarised in Table \ref{tab:ALMA_obs}.

The fields of view of the observations are shown in Figure \ref{fig:hst_alma}.
The data were calibrated and flagged in the standard manner using the Common Astronomy Software Applications package ({\tt CASA}; \citealt{CASA2022}).
The four tunings for band 3 and the three tunings for band 4 were separately combined to create individual continuum images for each band using {\tt CASA/tclean} task with the natural weighting.
The spectral channels where the emission line is detected were not used for the continuum imaging.
Each tuning data was also imaged to produce a cube with a channel width of 40 MHz ($\sim \SI{100}{km.s^{-1}}$) to search for emission lines after the continuum subtraction with {\tt CASA/uvcontsub} task.
The emission is cleaned down to $2\sigma$ level. For all images and cube data, the primary beam correction is applied.
We create images and cubes with a pixel size of 1\arcsec for both band3 and 4 in the M0417. In the R0032 region, a pixel size of 0.2\arcsec is used for both Band 3 and 4.

\begin{deluxetable*}{lcccccccc}
\tabletypesize{\scriptsize}
\tablewidth{0pt} 
\tablecaption{ALMA band 3 and 4 observation parameters \label{tab:ALMA_obs}}
\tablehead{
\colhead{Project code} & Band & \colhead{Target} & \colhead{Date}& \colhead{Antenna\tablenotemark{$a$}} & \colhead{PWV} & \colhead{Frequency range} & \colhead{Beam size} & \colhead{Cont. sensitivity} \\
\colhead{} & \colhead{} & \colhead{}& \colhead{} & \colhead{} & \colhead{(mm)} & \colhead{(GHz)} & \colhead{} & \colhead{($\mu$Jy/beam)}
} 
\startdata 
 & 3 & M0417-z365.ID46/.ID58/.ID121 & 2019/12 & C43-2 & 4.9--6.8 & 85.05--112.18 & $3.84\arcsec$$\times$$2.95\arcsec$ & 4.6 \\ 
2019.1.00949.S &  & (2-pointing mosaic) &  & (46--48) & & (T1--T4) & & \\
(PI: K. Kohno)& 4 & M0417-z365.ID46 & 2019/12 & C43-1/2 & 2.5--3.3 & 130.57--153.95\tablenotemark{$b$} & $2.66\arcsec$$\times$$1.96\arcsec$ & 7.6 \\
 &  & (1-pointing) &  & (42--43) & & (T5--T7) & & \\
\tableline
 & 3 & R0032-z239.ID127/.ID131/.ID198 & 2021/12 & C43-4/5 & 1.5--7.5 & 85.06--112.18 & $1.04\arcsec$$\times$$0.87\arcsec$ & 3.8 \\ 
 & & R0032-z299.ID208/.ID281/.ID304 & $\sim$2022/1 & (41--45) & & (T1--T4) & & \\
2021.1.01246.S &  & (2-pointing mosaic) &  &  & & & & \\
(PI: K. Kohno)& 4 & R0032-z239.ID127/.ID131 & 2022/8 & C43-5 & 0.3--0.9 & 130.56--153.94\tablenotemark{$b$} & $0.77\arcsec$$\times$$0.53\arcsec$ & 6.8 \\
 &  & R0032-z299.ID208 &  & (44--47) & & (T5--T7) & & \\
 &  & (2-pointing mosaic) &  & & & & & \\
\enddata
\tablecomments{a. Antenna configuration (the number in parentheses indicates the number of antennas)  b. There is a frequency gap of about 2.75 GHz in this range as shown in Figure \ref{fig:spectrum}.)}
\end{deluxetable*}

\subsection{JCMT SCUBA-2 450 \& 850 $\mu$m} 
We observed the M0417 cluster region at $\SI{450}{\mu m}$ and $\SI{850}{\mu m}$ 
with James Clerk Maxwell Telescope (JCMT)/Submillimetre Common-User Bolometer Array 2 (SCUBA-2) (Program ID: M20AP036, PI: K.~Kohno).
Observations were conducted on 2020 February 4, 16, July 8, and 31 using the CV Daisy scan, which creates a $\sim6\arcmin$ diameter circular map with a deep and uniform exposure coverage in the central $\sim1.5'$ diameter region. 
The total on-source time was 7.2 hr under excellent weather conditions, where the measured atmospheric opacities at 225 GHz were less than 0.05.
The typical beam sizes are 7$\arcsec$ and 15$\arcsec$ at 450~$\mu$m and 850~$\mu$m, respectively.

We reduced the data using the Dynamic Iterative Map Maker (DIMM) within the STARLINK Sub-Millimetre User Reduction Facility (SMURF) software package \citep{Chapin2013}.
The 
maps for each individual exposure were co-added and combined, using the STARLINK Pipeline for Combining and Analyzing Reduced Data (PICARD), and beam-match filtered. 
The resulting 1$\sigma$ noise levels were $\sim$3.9 mJy and $\sim$0.6 mJy at 450 $\mu$m and 850 $\mu$m bands, respectively.
The ALMA observations conducted by the ALCS did not detect any contamination sources near the positions of M0417-z365.ID46, ID58, and ID121. Therefore, no deblending was required, and the peak values at the ALMA Band 6 continuum positions were extracted as the source photometries.

\subsection{$K_{\rm S}$-band}
In order to obtain better constraints on the stellar properties of faint ALMA sources without any counterpart in $H$-band, we observed M0417 cluster region at $\SI{2.16}{\mu m}$ $Ks$-band with the \SI{6.5}{m} Magellan Baade Telescope/FourStar near-Infrared Camera on 2021 November 20 (PI: Wei-Hao Wang). 
We expect a significantly improved characterization of the stellar population (i.e., stellar mass, stellar age, visual extinction, etc.) by catching the 4000 \AA \ break between $H$- and $K_{\rm S}$-bands at $z\sim3.7$ (the redshift of the triple-lensed images in the M0417 region, as we will conclude in Section \ref{sec:specz}).
We obtained approximately 4.8 hr of on-source exposure. 
The final averaged seeing measured with \texttt{PSFEx} software is $\sim$0.64\arcsec.
We extracted photometries using a $D=1.5\arcsec$ aperture centered on the ALMA band 6 continuum positions for our target sources. We confirmed that this aperture maximizes the SNR of the enclosed flux density by drawing a growth curve.
The noise level is obtained by 1000 random apertures on the sigma-clipped map. We note that the values in the weight map do not deviate significantly from the average at the source location.
The aperture correction factor, calculated with the obtained PSF, is $\sim1.4$. 
The $5\sigma$-depth is measured to be 24.18 mag with the $D=1.5\arcsec$ aperture.

For the R0032 region, we make use of $K_S$-band imaging data taken with the High Acuity Wide field K-band Imager (HAWK-I) on Very Large Telescope. 
The Ks-band imaging of R0032 region (Program ID: 0103.A-0871(B), PI: A. Edge) is publicly available on the ESO Science archive.
The final averaged seeing is $\sim$1.12\arcsec, and the $5\sigma$-depth is 22.81 mag with a $D=2.0\arcsec$ aperture. No significant detection is obtained for our sources.


\section{Data analysis}\label{sec:results}
\subsection{Spectroscopic redshift identifications}\label{sec:specz}
\subsubsection{M0417-z365.ID46/.ID58/.ID121} \label{sec:M0417-z365}

Due to the sufficiently large beam size, the detection can be considered as a point source. 
We first created integrated-intensity maps around 99.1 GHz for Band 3 and around 148.6 GHz for Band 4, where bright emission lines were detected.
On these maps, we selected the pixels with peak values at each position of the triply lensed image. The spectra were then extracted from the cube image data at these specific pixels.

\begin{figure*}[tbh]
 \begin{center}
  \includegraphics[width=0.85\linewidth]{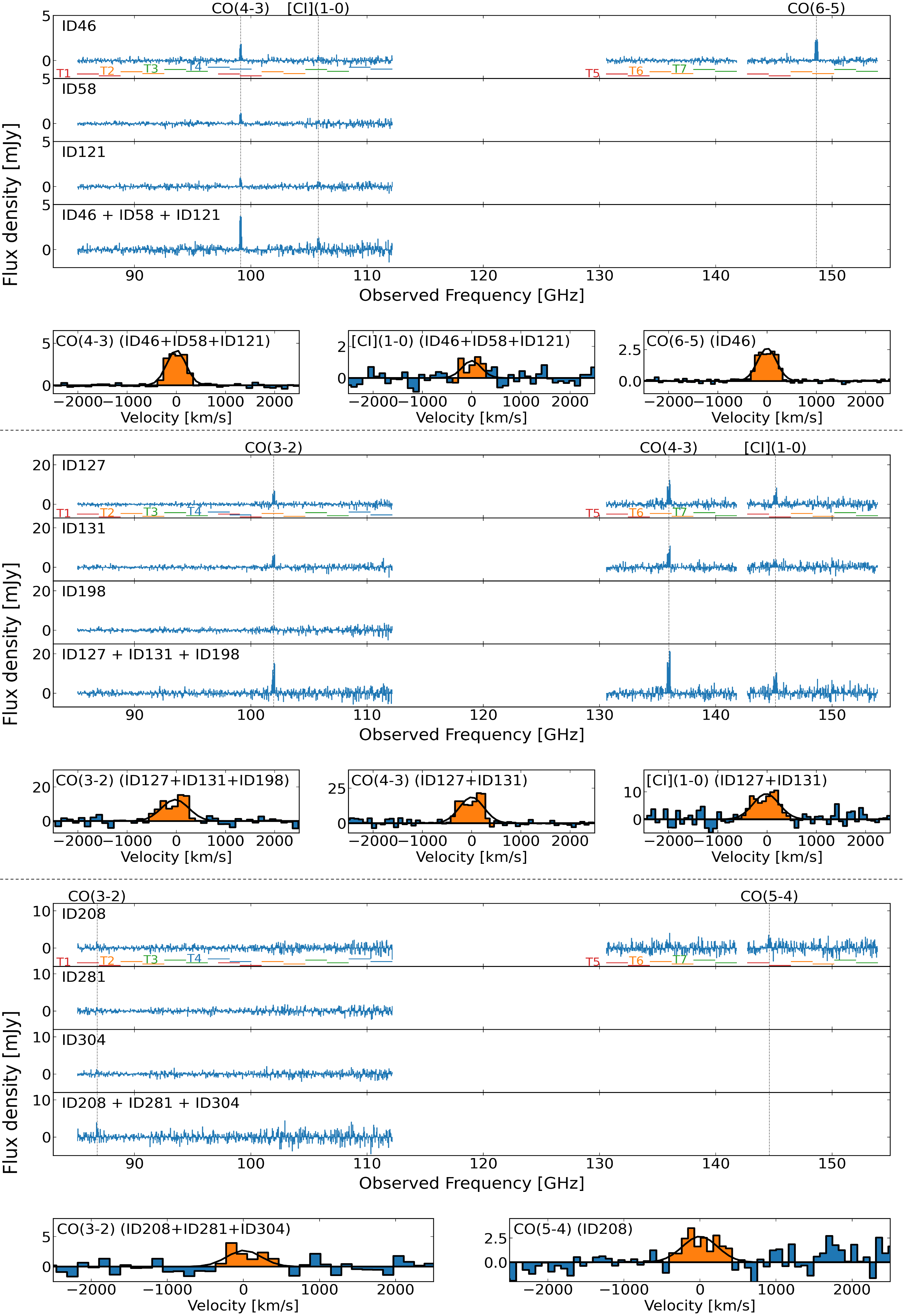}
 \end{center}
\vspace{-10pt}
\caption{The obtained spectra from ALMA line scan observations toward M0417 ID46, 58, and 121 (Top), R0032 ID127, 131, and 198 (Middle), and R0032 ID208, 281, and 304 (Bottom). For each triply lensed system, the top panel shows the full band-3/4 spectra from seven 7.5 GHz-wide independent tunings (T1-T7). The combined spectrum is also shown. The horizontal bars show the coverage of the seven tunings each of which has four spectral windows. The positions of the identified emission lines are shown with vertical black dashed lines. In the bottom panel, enlarged views of combined line spectra are shown. Velocities are relative to $z=3.652$, 2.391, and 2.985, respectively. The black solid lines are the best-fit single Gaussian. The orange shade indicates the integration range for the velocity-integrated flux.}
 \label{fig:spectrum}
\end{figure*}

Emission lines are clearly detected at $99.114\pm0.005\,\si{GHz}$ with peak flux densities of $8.6\sigma$, $7.3\sigma$, and $5.4\sigma$ from ID46, 58, and 121, respectively. 
Here, the noise level is calculated as the standard deviation of the line-free channel fluxes at the same pixels, specifically those within 50 channels of the emission line, with an absolute flux calibration error of 5\%.
We fitted the emission line with a single Gaussian and calculated the total flux as the integral over a frequency range of $1.5\times\text{FWHM}$.
For a more robust measurement, the FWHM of all lines is fixed to that of the CO(4-3) line with the highest SNR.
The total fluxes are $14.6\sigma$, $12.5\sigma$, and $8.9\sigma$ from ID46, 58, and 121, respectively. 
This line detection confirms the initial prediction from our lens model that these three sources are triply lensed images.
In addition, a bright emission line is detected at $148.650\pm0.006\,\si{GHz}$ from ID46 with a peak (total) $\text{flux}=9.6\sigma\ (22.3\sigma)$.
The noise level of the total flux is calculated by summing, in quadrature, the RMS noise from each channel derived from signal-free regions of the map across the channels containing the line.
Furthermore, when combining the spectra of the triply lensed images, a faint emission line is marginally detected at $105.84\pm0.02\,\si{GHz}$ with a total flux of $4.6\sigma$, although the peak flux density is only $3.4\sigma$.
We search the redshift solution that matches the three detected line combinations using a publicly-available code\footnote{\url{https://github.com/tjlcbakx/redshift-search-graphs}} \citep{Bakx2022}.
Since the frequency spacing between CO rotational transition lines is constant, detecting two CO lines can result in multiple possible redshift solutions. This code visually identifies the correct redshift solution by considering the observed frequencies, as well as the presence of detected lines and the absence of others.
The only solution is at $z=3.6517\pm0.0002$ and the three lines are CO(4-3), [C I](1-0), and CO(6-5).
This result is somewhat lower than the initial prediction from our lens model, $z=4.5\pm0.5$.
Figure \ref{fig:spectrum} shows the full ALMA band-3/4 spectra for the three triply lensed NIR-dark sources, including the other two triply lensed images in the R0032 region discussed below.
The line properties are summarized in Table \ref{tab:line_property}.
We similarly measured the band 3 and 4 continuum fluxes by extracting peak values near the source positions.


\subsubsection{R0032-z239.ID127/.ID131/.ID198} \label{sec:R0032-z239}
We used a circular aperture with a diameter of $D=4\arcsec$ centered on the ALCS band 6 continuum positions to extract spectra and continuum fluxes.
By drawing a growth curve, we confirmed that this aperture is sufficient to include source fluxes.
Emission lines are clearly detected at $101.963\pm0.008\,\si{GHz}$ with peak (total) fluxes of $8.8\sigma\ (13.6\sigma)$, $6.9\sigma\ (11.7\sigma)$, and $2.5\sigma\ (3.3\sigma)$ from ID127, 131, and 198, respectively. 
Thus, these three sources are triply lensed images as predicted from our lens model.
The total line fluxes are measured in the same manner as section \ref{sec:M0417-z365}.
In addition, bright emission lines are detected at $135.956\pm0.008\,\si{GHz}$ and $145.123\pm0.018\,\si{GHz}$ from ID127 and ID131.
From these lines, we conclude that this triply lensed source is at $z=2.39112\pm0.00019$ and detected emission lines are CO(3-2), CO(4-3), and [C I](1-0).
This result is consistent with the initial prediction from our lens model, $z=2.0\pm0.5$.

\subsubsection{R0032-z299.ID208/.ID281/.ID304}
We reconstructed the initial lens model of the R0032 cluster with the additional constraint that the R0032-z239.ID127/.ID131/.ID198 are triply lensed images and then found that it is highly likely that the R0032-z299.ID208/.ID281/.ID304 are also triply lensed images.
The spectra and continuum fluxes are measured in the same manner as in section \ref{sec:R0032-z239}.
A emission line is barely detected at $144.56\pm0.03\,\si{GHz}$ with peak (total) fluxes of $3.0\sigma\ (5.2\sigma)$ from ID208.
In addition, when combining the spectra of the triply lensed images, a faint emission line is detected at $86.77\pm0.02\,\si{GHz}$ with a peak (total) flux of $5.1\sigma\ (6.2\sigma)$.
These results conclude that this lensed source is at $z=2.9853\pm0.0009$ and detected emission lines are CO(3-2) and CO(5-4).
This result is consistent with the initial prediction from our lens model, $z=3.0\pm0.5$.
These consistencies enhance the reliability of the identified spectroscopic redshift, even though the line detections are marginal.
We emphasize that the detection of CO(3-2) line was made possible by the total magnification factor of $\sim15$, making the observation more than 100 times more efficient in terms of observation time.

\begin{deluxetable}{ccccc}
\tabletypesize{\scriptsize}
\tablewidth{0pt} 
\tablecaption{Velocity-integrated line flux density  \label{tab:line_property}}
\tablehead{
M0417-z365. & \colhead{ID46}& \colhead{ID58} & \colhead{ID121} & \colhead{Total}
} 
\startdata 
    $\mu$ & $5.1\substack{+2.6 \\ -2.6}$ & $3.9\substack{+1.6 \\ -1.6}$ & $3.4\substack{+1.4 \\ -1.4}$ & $12.5\substack{+3.3 \\ -3.3}$ \\
    CO(4-3)        & $0.88\pm0.06$        & $0.59\pm0.05$        & $0.48\pm0.05$        & $1.95\pm0.10$        \\
    \text{[C I](1-0)}       &    $0.16\pm0.06$     & $0.17\pm0.06$      & $0.16\pm0.07$        & $0.49\pm0.11$        \\ 
    CO(6-5)         & $1.20\pm0.05$        & --        & --        & $1.20\pm0.05$       \\
\hline\hline
R0032-z239. & \colhead{ID127}& \colhead{ID131} & \colhead{ID198} & \colhead{Total} \\
\tableline
    $\mu$ & $6.9\substack{+3.6 \\ -3.5}$ & $5.3\substack{+2.7 \\ -2.7}$ & $2.5\substack{+1.1 \\ -1.0}$ & $14.7\substack{+4.6 \\ -4.6}$ \\
    CO(3-2)     & $3.5 \pm 0.3$       & $3.7 \pm 0.3$        & $1.1\pm0.3$        & $8.3\pm0.5$        \\
    CO(4-3)   &  $5.9\pm0.3$     & $4.8\pm0.4$      & --        &  $10.7\pm0.5$         \\
    \text{[C I](1-0)}  & $3.7\pm0.4$        & $2.2\pm0.5$        & --        & $5.9\pm0.6$        \\
\hline\hline
R0032-z299. & \colhead{ID208}& \colhead{ID281} & \colhead{ID304} & \colhead{Total} \\
\tableline
    $\mu$ & $7.9\substack{+4.1 \\ -4.0}$ & $4.3\substack{+1.8 \\ -1.8}$ & 3.0$\substack{+1.3 \\ -1.2}$ & $15.2\substack{+4.6 \\ -4.6}$\\
    CO(3-2)        & $0.57\pm0.17$        & $0.42\pm0.15$        & $0.59\pm0.13$        & $1.6\pm0.2$        \\
    CO(5-4)  &    $1.6\pm0.3$     & --      & --        & $1.6\pm0.3$    \\
\enddata
\tablecomments{$\mu$ is the magnification factor. The velocity-integrated line flux density is in unit of \si{Jy.km.s^{-1}}. The magnification factor is not corrected.}
\end{deluxetable}

\subsection{Mass model}\label{sec:lensmodel}
We utilize the publicly available {\tt glafic} code \citep{Oguri2010, Oguri2021} to construct our fiducial lens models, following \cite{Kawamata2016} and \cite{Okabe2020}.
The multiple images that constrain the lens model are sourced from \cite{Mahler2019} for the M0417 cluster and from \cite{Acebron2020} for the R0032 cluster.
Additionally, the three triply lensed systems identified in this study are included to provide further constraints.
To assess the statistical uncertainty of the magnification factor, we employ an MCMC method, taking into account uncertainties in the positions of the multiple images ($\Delta x= \Delta y=0.4\arcsec$) and photometric redshifts ($\Delta z= 0.6$).
The median value is chosen as the representative value, with the 16th and 84th percentiles serving as the uncertainty range.
To account for potential systematic errors in the lens model, we introduce an additional error based on a comparison with other lens models that are constructed with different software, as detailed by \cite{Fujimoto2024ApJS..275...36F}.
Specifically, we incorporate an error of 40\% for magnifications below 5 and 50\% for those above 5.
The obtained magnification factors can be found in Table \ref{tab:line_property}.

\subsection{SED fitting}
We use the {\tt CIGALE} code \citep{Boquien2019} to perform SED fitting and estimate key physical parameters.
The photometries used for the SED fitting are listed in Table \ref{tab:photometry}.
{\tt CIGALE} models the SED panchromatically based on an energy balance principle, where all the absorbed energy is then re-emitted by the dust in the mid/far-infrared domains.
Following \cite{Huang2023}, instead of the default fixed grid parameter search in {\tt CIGALE}, we implement a Bayesian-based approach using {\tt DYNESTY} package \citep{Speagle2020}.
This allows for efficient exploration of extensive parameter spaces and computation of the Bayesian posterior distribution for the free parameters.
We use flat prior distributions for all parameters.

We assume a nonparametric star-formation history (SFH) \citep{Leja2019} and adopt the stellar population model from \cite{Bruzual2003} with solar metallicity. 
The dust attenuation law is taken from \cite{Calzetti2000}.
For the dust infrared emission model, we use \cite{Draine2014} templates.
We note that bright optical emission lines such as H$\alpha$ and [O III]$_{5007}$ don't fall into the photometric bands used for our sources at their redshifts.
For each multiple image, we performed SED fitting and, based on the posterior distribution, determined the median value as the best-fit parameter, accompanied by the 16--84\% credible interval. 
In our case, differential magnification is negligible, meaning that the physical quantities scale linearly with the magnification factor. Therefore, the intrinsic physical quantities are derived by dividing the obtained apparent physical quantities by the magnification factor.
After correcting each value for magnification, we took the intrinsic physical quantity as the average weighted by the inverse square of the uncertainties.
The best-fit SED models along with their 16--84\% credible interval are shown in Figure \ref{fig:sed}.
The obtained intrinsic physical parameters along with their 16--84\% percent credible interval are listed in Table \ref{tab:all_params}.
For the fiducial SFR value, we take one that is averaged over 100 Myr.
The SFR can also be estimated from the total infrared luminosity ($L_{\mathrm{IR}}$) using the following relation:
$\mathrm{SFR} = 4.5 \times 10^{-44} \, \mathrm{M_\odot \, yr^{-1} \, (erg \, s^{-1})^{-1}} \, L_{\mathrm{IR}}$,
which is a calibration by \cite{Madau2014}, but we multiply a factor of 0.63 to account for the correction from the Salpeter IMF to the Chabrier IMF. Using this equation, the SFR values inferred from $L_{\mathrm{IR}}$ are $167_{-69}^{+63}~\mathrm{M_\odot \, yr^{-1}}$ for M0417-z365, $92_{-35}^{+35}~\mathrm{M_\odot \, yr^{-1}}$ for R0032-z239, and $33_{-16}^{+12}~\mathrm{M_\odot \, yr^{-1}}$ for R0032-z299. These values are consistent within the uncertainties with the SFRs derived from the SED fitting.

\begin{figure}[htb]
\begin{center}
\includegraphics[width=\linewidth]{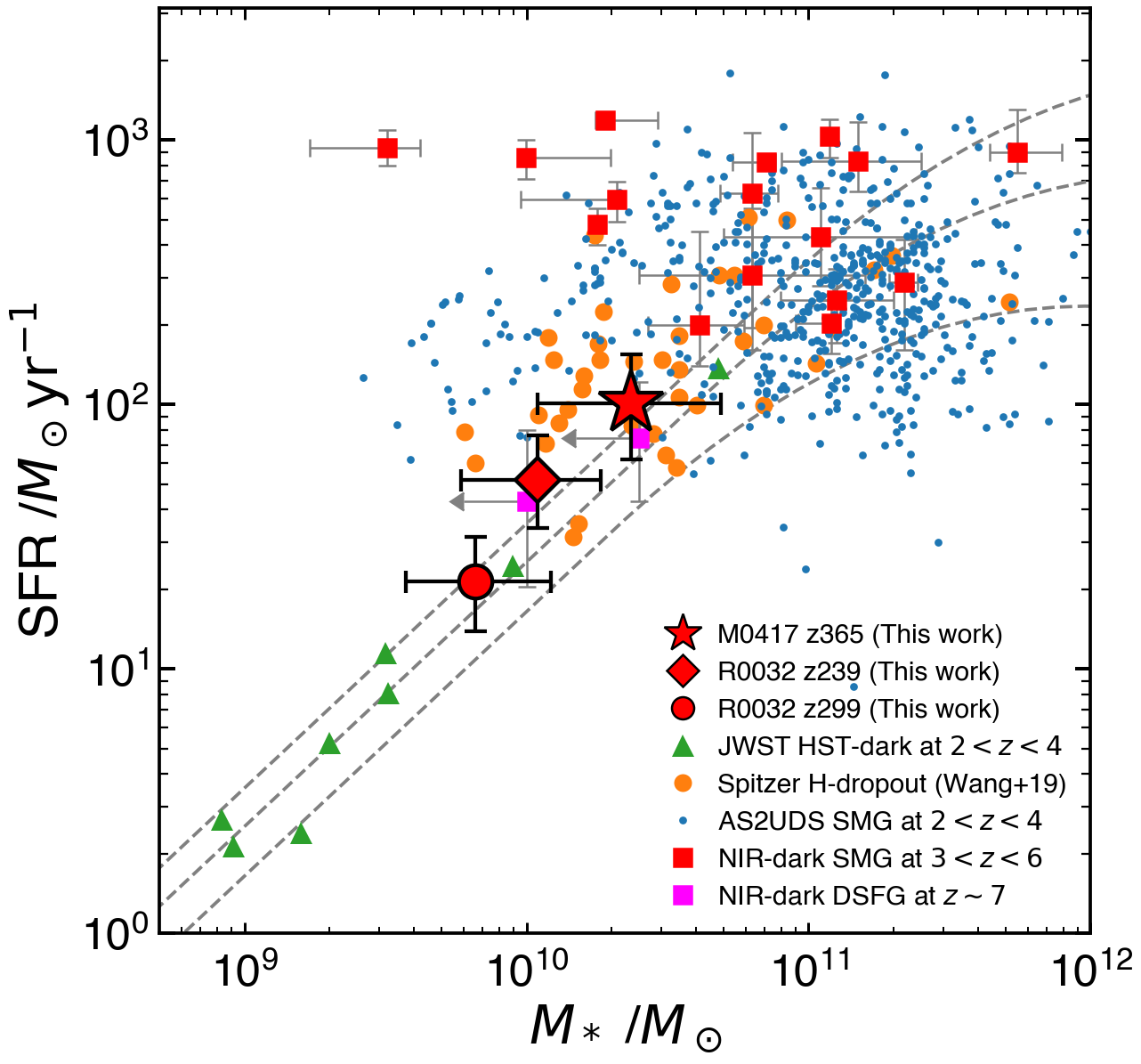}
\end{center}
\caption{SFR against stellar mass relation for our three intrinsic sources (red star/diamond/circle) is shown along with other populations for comparison. The green triangle represents {\it JWST}/NIRCam selected $H$-dropouts at $2<z<4$ \citep{Barrufet2023}. The orange circle shows ALMA and {\it Spitzer}/IRAC-detected $H$-dropouts \citep{Wang2019Nat}. The blue point represents $2<z<4$ SMGs from the AS2UDS samples \citep{Dudzeviciute2020}. The red squares are NIR-dark SMGs from the literature \citep{Casey2019, Williams2019, Zhou2020, Umehata2020, Riechers2020, Manning2022, Smail2023, Kokorev2023, Sun2024}. Two magenta squares are NIR-dark DSFGs at $z\sim7$ \citep{Fudamoto2017}. 
The gray dashed lines indicate the star-forming main sequence at $z=3$, 4, and 5 \citep{Schreiber2015}.
\label{fig:mstar_sfr}}
\end{figure}

Figure \ref{fig:mstar_sfr} shows the SFR versus $M_*$ relation of our three intrinsic sources. 
For comparison, {\it JWST}/NIRCam (long-wavelength filter)-selected $H$-band dropouts ($3\sigma$ depth of $H>\SI{27}{mag}$, \citealt{Barrufet2023}), ALMA and {\it Spitzer}/IRAC-detected $H$-band dropouts (typical $5\sigma$ depth of $H>\SI{27}{mag}$, \citealt{Wang2019Nat}), $2<z<4$ SMGs from the AS2UDS samples \citep{Dudzeviciute2020}, NIR-dark SMGs \citep{Casey2019, Williams2019, Zhou2020, Umehata2020, Riechers2020, Manning2022, Smail2023, Kokorev2023, Sun2024}, and NIR-dark DSFGs \citep{Fudamoto2021} are also shown.
Some of these previous studies estimate physical quantities using {\tt MAGPHYS} \citep{dacunha2015}. Different SED fitting codes often produce systematic differences in results due to various assumptions they employ, especially the dust attenuation models. 
\cite{Uematsu2024} investigated whether there are systematic differences in physical quantities obtained with {\tt MAGPHYS} and {\tt CIGALE} with the Calzetti attenuation curve, using ALCS sample.
They found that while the SFR estimates agree well, the stellar mass estimates from {\tt CIGALE} are 0.37 dex lower at $M_*=10^{10.5} M_\odot$. 
In any case, our sample remains significantly lower in SFR and stellar mass compared to the SMGs in previous studies.
Our three samples lie on the main sequence (offset from the main sequence $\Delta \mathrm{MS}=\mathrm{SFR}/\mathrm{SFR_{MS}}$ is within a factor of 3), exhibiting a $M_*$--SFR relation similar to typical star-forming galaxies selected in NIR bands (e.g., \citealt{Schreiber2015, Straatman2016}).
\cite{Wang2019Nat} and \cite{Xiao2023} suggest that such dust-obscured galaxies on the main sequence are a dominant population of the massive galaxies ($M_*\gtrsim10^{9.5}\,M_\odot$) at $3<z<6$.
Among our samples, M0417-z365 is located at a region where $H$-dropouts of \cite{Wang2019Nat} are sampled. 
R0032-z299 is positioned on the lower mass and SFR region, bridging the gap with the dust-obscured galaxy candidates recently explored by {\it JWST}.

\movetabledown=40mm
\begin{rotatetable*}
\begin{deluxetable*}{lcc|ccc|ccc|ccc|c}
\tablecolumns{13}
\tabletypesize{\scriptsize}
\tablewidth{0pt} 
\tablecaption{Photometries of the three NIR-dark triply lensed systems (the magnification factor is not corrected) \label{tab:photometry}}
\tablehead{
\colhead{} & \colhead{} & \colhead{} & \multicolumn{3}{c}{M0417-z365} & \multicolumn{3}{c}{R0032-z239} & \multicolumn{3}{c}{R0032-z299} & \colhead{}\\
\colhead{Band} & \colhead{Wavelength} & \colhead{Unit} & \colhead{ID46} & \colhead{ID58} & \colhead{ID121} & \colhead{ID127} & \colhead{ID131} & \colhead{ID198} & \colhead{ID208} & \colhead{ID281} & \colhead{ID304} & \colhead{Reference}
} 
\startdata 
    {\it HST}-F606W & \SI{606}{nm} & nJy & blended & $<26$ & $<26$ & $<53$ & $<53$ & $<53$ & $<53$ & $<53$ & $<53$ & K22 \\
    {\it HST}-F814W & \SI{814}{nm} & nJy & blended & $<68$ & $<68$ & $<63$ & $<63$ & $<63$ & $<63$ & $<63$ & $<63$ & K22 \\
    {\it HST}-F105W & \SI{1.05}{\mu m} & nJy & blended & $<106$ & $<106$ & $<83$ & $<83$ & $<83$ & $<83$ & $<83$ & $<83$ & K22 \\
    {\it HST}-F125W & \SI{1.25}{\mu m} & nJy & blended & $<135$ & $<135$ & $<137$ & $<137$ & $<137$ & $<137$ & $<137$ & $<137$ & K22 \\
    {\it HST}-F140W & \SI{1.40}{\mu m} & nJy & blended & $<116$ & $<116$ & $207\pm31$ & $<119$ & $<119$ & $<119$ & $<119$ & $<119$ & K22 \\
    {\it HST}-F160W & \SI{1.60}{\mu m} & nJy & blended & $<111$ & $<111$ & $262\pm23$ & $707\pm121$ & $362\pm42$ & $<88$ & $<88$ & $<88$ & K22 \\
    FourStar-Ks & \SI{2.2}{\mu m} & \si{\mu Jy} & blended & $0.85\pm0.22$ & $<0.63$ & $\cdots$ & $\cdots$ & $\cdots$ & $\cdots$ & $\cdots$ & $\cdots$ & This work \\
    HAWKI-Ks & \SI{2.2}{\mu m} & \si{\mu Jy} & $\cdots$ & $\cdots$ & $\cdots$ & $<2.4$ & $<2.4$ & $<2.4$ & $<2.4$ & $<2.4$ & $<2.4$ & This work \\
    IRAC-ch1 & \SI{3.6}{\mu m} & \si{\mu Jy} & blended & $1.59\pm0.08$ & $1.33\pm0.07$ & $4.77\pm0.19$ & $11.5\pm0.2$ & $3.77\pm0.15$ & $2.02\pm0.19$ & $2.5\pm0.2$ & $2.65\pm0.16$ & K22 \\
    IRAC-ch2 & \SI{4.5}{\mu m} & \si{\mu Jy} & blended & $2.85\pm0.14$ & $2.67\pm0.14$ & $8.32\pm0.17$ & $17.85\pm0.17$ & $6.44\pm0.14$ & $3.22\pm0.17$ & $2.5\pm0.3$ & $4.10\pm0.14$ & K22 \\
    SPIRE    & \SI{250}{\mu m} & mJy & $45.4\pm5.8$ & $11.1\pm7.3$ & $10.9\pm6.3$ & $13.2\pm5.3$ & $12.7\pm5.3$ & $33.8\pm7.7$ & $<15.6$ & $<15.7$ & $\cdots$ & S22 \\
    SPIRE    & \SI{350}{\mu m} & mJy & $39.0\pm6.2$ & $14.8\pm5.3$ & $6.9\pm4.7$  & $30.2\pm7.4$ & $15.6\pm7.4$ & $24.9\pm8.4$ & $<21.7$ & $<21.8$ & $<27.0$ & S22 \\
    SCUBA-2  & \SI{450}{\mu m} & mJy & $24.9\pm4.0$ & $14.3\pm4.0$ & $13.8\pm3.8$ & $\cdots$ & $\cdots$ & $\cdots$ & $\cdots$ & $\cdots$ & $\cdots$ & This work \\
    SPIRE    & \SI{500}{\mu m} & mJy & $30.7\pm5.1$ & $11.1\pm5.7$ & $17.6\pm5.2$ & $<38.9$ & $32.7\pm11.4$ & $<23.0$ & $10.6\pm8.8$ & $<21.8$ & $<22.7$ & S22 \\
    SCUBA-2  & \SI{850}{\mu m} & mJy & $11.09\pm0.65$ & $5.4\pm0.66$ & $5.28\pm0.65$ & $\cdots$ & $\cdots$ & $\cdots$ & $\cdots$ & $\cdots$ & $\cdots$ & This work \\
    ALMA-B6  & \SI{1.2}{mm}    & mJy & $4.3\pm0.5$ & $3.1\pm0.4$ & $2.1\pm0.3$ & $4.5\pm0.5$ & $3.9\pm0.4$ & $1.4\pm0.2$ & $1.2\pm0.2$ & $1.2\pm0.2$ & $1.1\pm0.2$ & This work\\
    ALMA-B4  & \SI{2.1}{mm}    & mJy & $0.395\pm0.018$ & $\cdots$ & $\cdots$ & $1.1\pm0.1$ & $0.88\pm0.09$ & $\cdots$ & $0.41\pm0.08$ & $\cdots$ & $\cdots$ & This work\\
    ALMA-B3  & \SI{3.0}{mm}    & mJy & $0.095\pm0.007$ & $0.051\pm0.005$ & $0.039\pm0.007$ & $0.25\pm0.04$ & $0.23\pm0.04$ & $<0.09$ & $<0.08$ & $<0.08$ & $0.09\pm0.03$ & This work\\
\enddata
\end{deluxetable*}
\end{rotatetable*}

\subsection{Molecular gas properties}
We estimate the total molecular gas mass $M_{\mathrm{gas}}$ using CO(3-2) line for the R0032-z239 and R0032-z299 and CO(4-3) line for the M0417-z365, respectively.
We assume an average CO excitation of $r_{32/10}=0.58\pm0.10$ and $r_{43/10}=0.30\pm0.08$ \citep{Boogaard2020}, where $r$ represents brightness temperature ratio. 
The value of CO-to-H$_2$ conversion factor $\alpha_{\mathrm{CO}}$ depends on various properties including metallicity, galactic environment, and morphology. 
In the following, we will show both results obtained with the assumption of $\alpha_{\mathrm{CO}}=3.6\ \mathrm{M_\odot (K.km.s^{-1}.pc^2)^{-1}}$, which is often assumed for massive MS galaxies (e.g., \citealt{Aravena2019}), and of $\alpha_{\mathrm{CO}}=1.0\ \mathrm{M_\odot (K.km.s^{-1}.pc^2)^{-1}}$, which is often assumed for SMGs and compact starbursts (e.g., \citealt{Bothwell2013}).
We also calculate the molecular gas mass using [C I](1-0) luminosity for M0417-z365 and R0032-z239 following \cite{Umehata2020}.
The gas masses derived from both CO and [C I] luminosities are consistent provided that the $\alpha_{\mathrm{CO}}$ ranges between 1.0 and 3.6.
We use CO-based molecular gas mass as a fiducial value.
The gas depletion time $\tau_{\mathrm{dep}}=M_{\mathrm{gas}}/\mathrm{SFR}$ and gas fraction $f_{\mathrm{gas}}=M_{\mathrm{gas}}/M_*$ are also derived in conjunction with the SED fitting results.
The obtained values are listed in Table \ref{tab:all_params}.

\begin{figure*}[htb]
\begin{center}
    \includegraphics[width=1\linewidth]{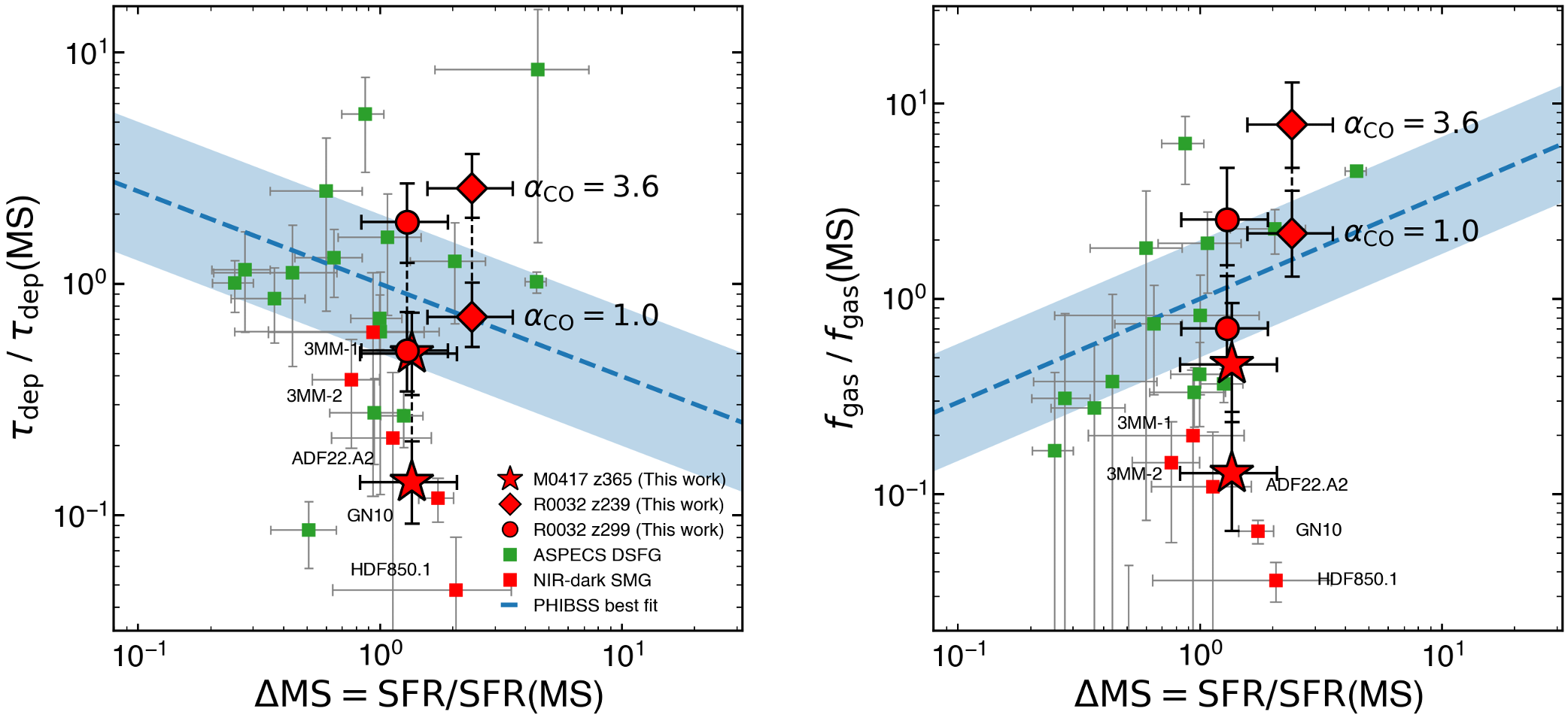}
\end{center}
\caption{Gas depletion time against $\Delta$MS (Left) and gas fraction against $\Delta$MS (Right). The dashed line and blue shaded region show the best-fit line for the PHIBBS samples ($\tau_{\mathrm{gas}}$/$\tau_{\mathrm{gas}}(\mathrm{MS})=\Delta\mathrm{MS}^{-0.40}$ and $f_{\mathrm{gas}}$/$f_{\mathrm{gas}}(\mathrm{MS})=\Delta\mathrm{MS}^{0.53}$) derived in \cite{Tacconi2018} and its $\pm0.3$ dex offset, respectively. The green square represents the ASPECS DSFG samples from \cite{Aravena2020}, and the red square shows the NIR-dark SMGs for which gas mass is measured based on emission lines (\citealt{Williams2019, Riechers2020, Umehata2020, Sun2024}).
\label{fig:dms_dep_fgas}}
\end{figure*}

The molecular gas depletion time ($\tau_{\mathrm{dep}}=M_{\mathrm{gas}}/\mathrm{SFR}$) and gas fraction ($f_{\mathrm{gas}}=M_{\mathrm{gas}}/M_*$) depend on redshift, stellar mass, and offset from the main sequence ($\Delta\mathrm{MS}$). 
\cite{Tacconi2018} compiled molecular gas mass data of star-forming galaxies at $0<z<4$, covering a wide range of stellar mass and $\Delta\mathrm{MS}$ and derived the scaling relations of $\tau_{\mathrm{dep}} \propto (1+z)^{-0.62}\times \Delta\mathrm{MS}^{-0.44}\times M_*^{0.09}$ and $f_{\mathrm{gas}} \propto (1+z)^{2.5}\times \Delta\mathrm{MS}^{0.52}\times M_*^{-0.36}$.
Figure \ref{fig:dms_dep_fgas} shows $\tau_{\mathrm{dep}}$ and $f_{\mathrm{gas}}$ as a function of $\Delta\mathrm{MS}$, normalized using the scaling relations.
For comparison, we also plot ASPECS DEFGs \citep{Aravena2020}, NIR-dark SMGs in literature \citep{Williams2019,Riechers2020,Umehata2020,Sun2024}.
The gas depletion time and gas fraction of R0032-z299 are consistent with the scaling relation regardless of $\alpha_{\mathrm{CO}}$ value. This suggests that this galaxy is a normal star-forming galaxy undergoing secular evolution.
M0417-z365 seems to exhibit a shorter $\tau_{\mathrm{dep}}$ and a smaller $f_{\mathrm{gas}}$, deviating by 0.80 dex and 0.96 dex from the scaling relations, respectively, assuming $\alpha_{\mathrm{CO}} = 1.0$ (0.25 dex and 0.41 dex if $\alpha_{\mathrm{CO}} = 3.6$).
For R0032-z239, these quantities are consistent with the scaling relation when $\alpha_{\mathrm{CO}}$ is close to 1.
Examining our results with those of previous literature suggests that the NIR-dark galaxy population has a wide range of molecular gas depletion time and gas fraction.
We will further discuss the interpretations of the molecular gas properties in section \ref{sec:discussion}.

Using the molecular gas masses derived above and the dust masses estimated from the SED fitting, we calculate the dust-to-gas mass ratio ($\delta_{\mathrm{GDR}} = M_{\mathrm{dust}} / M_{\mathrm{gas}}$). Assuming $\alpha_{\mathrm{CO}} = 1.0$, the dust-to-gas mass ratios based on CO luminosity are $23 \pm 13$ for M0417-z365, $31 \pm 18$ for R0032-z239, and $23 \pm 14$ for R0032-z299.
When compared with the sample of SMGs at $z = 1–5$ presented in \cite{Birkin2021}, where molecular gas masses are derived using CO with the assumption of $\alpha_{\mathrm{CO}} = 1.0$, the $\delta_{\mathrm{GDR}}$ values for our sources appear to be smaller, despite the large scatter in the sample. If the $\delta_{\mathrm{GDR}}$ of our sources is comparable to that of the $z=1-5$ SMG sample, this could imply that $\alpha_{\mathrm{CO}}$ for our sources is larger than 1.0.

\subsection{Morphology of star-forming region} \label{sec:morphology}
We reconstruct ALMA band 6 (\SI{1.2}{mm}) continuum images for the three triply lensed systems observed by ALCS using the {\tt glafic} version 2.
Ideally, it would be preferable to simultaneously fit the triply lensed images using a single source model.
However, galaxy cluster lenses have complex gravitational potential such as small-scale substructures that are not accounted for in the parametric mass model. 
As a result, the positions of these multiple images typically often deviate by about $\sim 0.5\arcsec$ (e.g., \citealt{Oguri2013}). 
In this study, we perform individual fittings for each of the three images and then adopt their average value.

First, we refine the lens model by including the multiple images identified in this study as constraints, in addition to those previously recognized.
Subsequently, we put a S\'{e}rsic source with a S\'{e}rsic index $n$ of 1, characterized by six free parameters: $xy$-coordinates, flux, effective radius (half-light radius), major-to-minor axis ratio, and position angle) on the source plane and all source parameters are optimized with the lens parameters hold constant.
For $\chi^2$ optimization, only the area within a 3\arcsec-radius circle centered on each source was considered.

\begin{figure}
    \centering
    \includegraphics[width=0.95\linewidth]{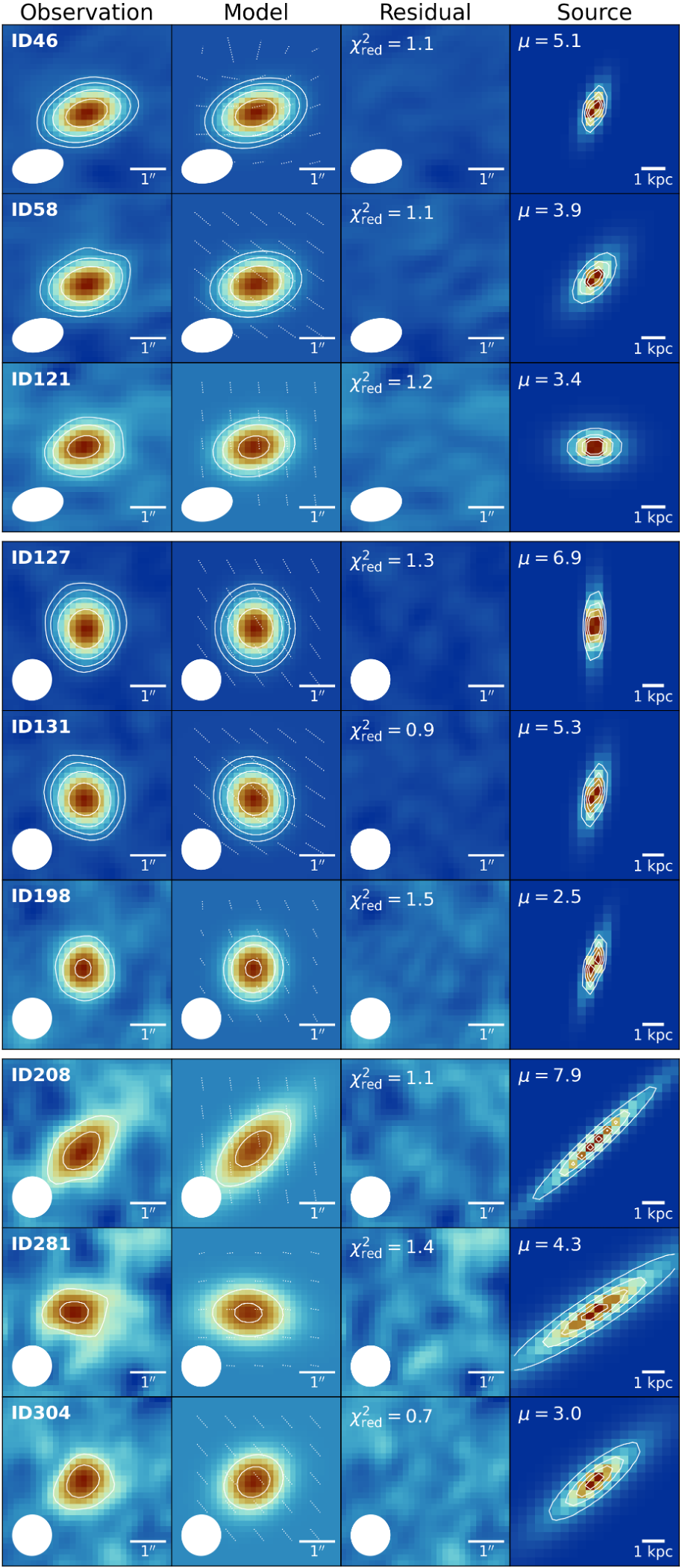} 
    \caption{Source reconstruction for the three triply lensed systems. From left to right, the observed ALCS band6 (\SI{1.2}{mm}) cleaned image, best-fit model, residual, and reconstructed source are shown. The white contours on the observed image and model image are drawn at [4, 8, 16, 32]$\times \sigma$. The peak residuals are less than $3\sigma$. The black contours in the source panel show 20\%, 40\%, 60\%, and 80\% of the peak. The synthesized beam size is displayed in the lower-left corner of each panel. The dotted line in the model panel shows the shear direction and strength at the position.
    The reduced chi-square values and magnification factors are annotated on the residual and source images, respectively.
    }
    \label{fig:glafic}
\end{figure}

The uncertainty in source reconstruction arises from both the lens model and the observed image.
To address the uncertainty of the lens model, we utilize the multitude of lens models generated and accepted via the MCMC method when constructing the lens model as described in Section \ref{sec:lensmodel}.
We repeat the same {\tt glafic} fitting process with these lens models. 
In addition, to consider the uncertainty of the observed image, we add a randomly generated noise map with the same noise level as the observed image to the best-fit (lensed) model image (which does not include noise; the image in the Model panel of Figure \ref{fig:glafic}) and iterate the fitting. 
Given the pixel-correlated noise characteristic of interferometric images, we use a noise map convolved with the dirty beam of the ALMA observation.
After conducting the fitting procedure 20,000 times in total, we adopt the median value as the best-fit value and use the 16--84\% percent interval as the uncertainty.
In Appendix \ref{appe:resolution}, we further discuss the feasibility of spatially resolving the dust continuum from our ALMA observations.

\begin{figure}[h]
 \begin{center}
  \includegraphics[width=\linewidth]{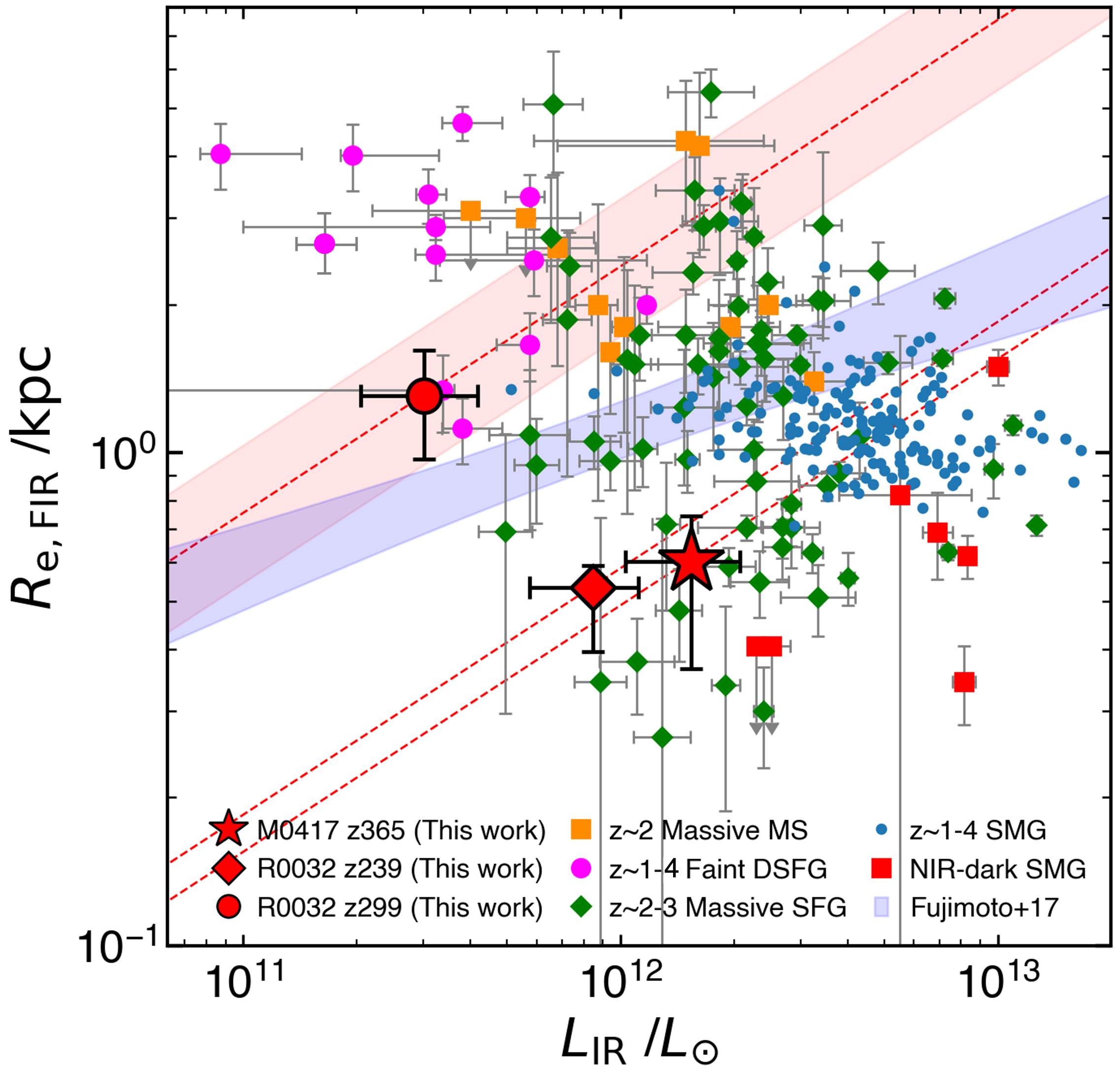}
 \end{center}
 \caption{Intrinsic infrared luminosities versus circularized effective radii for our three samples, massive main sequence galaxies at $z\sim2$ \citep{Rujopakarn2016}, faint DSFGs at $z\sim1-4$ \citep{Gonzalez-Lopez2017, Laporte2017b, Pope2023ApJ...951L..46P, Mizener2024ApJ...970...30M}, massive ($M_*>10^{11}\,M_\odot$) SFGs at $z\sim2-3$ \citep{Tadaki2020}, AS2UDS SMGs at $z=1-4$ \citep{Gullberg2019, Dudzeviciute2020, Smail2021}, and NIR-dark SMGs \citep{Umehata2017, Umehata2020, Gomez2022}. The red dashed lines (with a shaded region of $1\sigma$ uncertainty for R0032-z299) correspond to the constant infrared surface density relations for our samples. The blue-shaded region shows the scaling relation with $1\sigma$ uncertainty derived from a statistical ALMA archival study \citep{Fujimoto2017}.}
 \label{fig:re_fir}
\end{figure}

Figure \ref{fig:glafic} shows the best-fit model for the three triply lensed systems.
The obtained intrinsic circularized effective radius $r_e$ and apparent axis ratio for them are listed in Table \ref{tab:all_params}.
Figure \ref{fig:re_fir} shows the relation of the intrinsic infrared luminosities and circularized effective radii for our three samples along with massive ($10^{10.5}\,M_\odot < M_*< 10^{11}\,M_\odot$) MS SFGs at $z\sim 2$ \citep{Rujopakarn2016}, faint DSFGs with SFR of 10--100 $M_\odot/\mathrm{yr^{-1}}$ at $z=1-4$ \citep{Gonzalez-Lopez2017, Laporte2017b, Pope2023ApJ...951L..46P, Mizener2024ApJ...970...30M}, massive ($M_*>10^{11}\,M_\odot$) DSFGs at $z\sim2$ \citep{Tadaki2020}, SCUBA-2 selected SMGs at $z=1-4$ \citep{Gullberg2019, Dudzeviciute2020}, and NIR-dark SMGs \citep{Umehata2017, Umehata2020, Gomez2022}.
While the infrared luminosity of M0417-z365 and R0032-z239 is much lower than that of SMGs, their dust-emitting regions are as compact as those of NIR-dark SMGs.
The tendency for compact objects to experience stronger dust attenuation has been suggested by other studies (e.g., \citealt{Smail2021, Gomez-Guijarro2023}).
On the other hand, R0032-z299, the faintest one in our sample, has a more extended dust continuum of $R_{\mathrm{e, 1.2mm}}=1.3\pm0.3$ kpc.
The infrared surface densities of M0417-z365, R0032-z239, and R0032-z299 are $6.6\substack{+7.8 \\ -3.3}\times 10^{11}$, $4.7\substack{+2.9 \\ -1.9}\times 10^{11}$, and $2.8\substack{+2.6 \\ -1.3}\times 10^{10}\,L_\odot\si{kpc^{-2}}$, respectively.
We will further discuss the interpretations of the size-luminosity relation in section \ref{sec:discussion}. 

\begin{deluxetable*}{llccc}
\tabletypesize{\scriptsize}
\tablewidth{0pt} 
\tablecaption{Derived intrinsic physical properties for our three samples \label{tab:all_params}}
\tablehead{
\multicolumn{2}{c}{Parameters} & \colhead{M0417-z365}& \colhead{R0032-z239} & \colhead{R0032-z299}  
} 
\startdata 
Redshift & $z$ & $3.6517\pm0.0002$ &$2.3911\pm0.0002$ &$2.9853\pm0.0009$ \\
Stellar mass & $\log_{10}(M_*/M_\odot)$ & $10.37\substack{+0.23 \\ -0.47}$ & $10.04\substack{+0.20 \\ -0.30}$ & $9.82\substack{+0.19 \\ -0.37}$\\
Infrared luminosity & $\log_{10}(L_{\mathrm{IR}}/L_\odot)$ & $12.19\substack{+0.14 \\ -0.15}$ & $11.93\substack{+0.14 \\ -0.14}$ & $11.48\substack{+0.14 \\ -0.17}$\\
Star formation rate (averaged over 100 Myr) & SFR [$M_\odot \si{yr^{-1}}$] & $101\substack{+39 \\ -54}$ & $52\substack{+18 \\ -24}$ & $21\substack{+8 \\ -10}$ \\
Dust attenuation & $A_V$ [mag] & $4.5\substack{+0.4 \\ -0.4}$ & $4.6\substack{+0.19 \\ -0.20}$ & $4.0\substack{+0.4 \\ -0.5}$\\
Dust mass & $\log_{10}(M_\mathrm{dust}/M_\odot)$ & $8.46\substack{+0.14 \\ -0.14}$ & $8.73\substack{+0.14 \\ -0.15}$ & $8.29\substack{+0.14 \\ -0.14}$\\
Offset from the main sequence & $\Delta \mathrm{MS}$ & $1.4\substack{+0.5 \\ -0.7}$ & $2.4\substack{+1.1 \\ -0.8}$ & $1.3\substack{+0.5 \\ -0.6}$ \\
Molecular gas mass (CO-based, $\alpha_{\mathrm{CO}}=1$) & $\log_{10}(M_\mathrm{gas}^{\mathrm{CO})}/M_\odot)$ &$9.83\pm0.14$ & $10.22\pm0.15$ & $9.66\pm0.16$\\
Molecular gas mass ([C I]-based) & $\log_{10}(M_\mathrm{gas}^{\mathrm{[C I]})}/M_\odot)$ &$9.92\pm0.15$ & $10.71\pm0.17$ & $\cdots$\\
Gas depletion time ($\alpha_{\mathrm{CO}}=1$) & $\tau_{\mathrm{gas}}$ [Gyr] & $0.07\substack{+0.02 \\ -0.03}$ & $0.35\substack{+0.09 \\ -0.14}$ & $0.22\substack{+0.07 \\ -0.10}$ \\
Gas fraction ($\alpha_{\mathrm{CO}}=1$) & $f_{\mathrm{gas}}\equiv M_\mathrm{gas}/M_*$ & $0.30\substack{+0.15 \\ -0.32}$ & $1.7\substack{+0.7 \\ -1.1}$ & $0.7\substack{+0.3 \\ -0.6}$ \\
1.2 mm continuum circularized radius & $R_{\mathrm{e, 1.2 mm}}$ [kpc] & $0.60\substack{+0.14 \\ -0.24}$ & $0.53\substack{+0.06 \\ -0.14}$ & $1.3\substack{+0.3 \\ -0.3}$ \\
Apparent axis ratio & $b/a$ & $0.50\substack{+0.27 \\ -0.32}$ & $0.24\substack{+0.04 \\ -0.07}$ & $0.17\substack{+0.11 \\ -0.07}$\\
\enddata
\end{deluxetable*}

\subsection{PDR properties}\label{sec:pdr} 
The photodissociation regions (PDRs) play a dominant role in the heating, cooling, ionization, and dissociation of molecules in the neutral interstellar medium (ISM) \citep{TH1985ApJ...291..722T, BT1994ApJ...427..822B}. 
The heating occurs when far-UV photons penetrate the ISM and are absorbed by Polycyclic Aromatic Hydrocarbons (PAHs). These particles release photoelectrons, which collisionally excite the gas, ultimately converting the energy into dust emission.
The gas cooling occurs primarily through FIR fine structure lines, such as [C II]$_158$, [O I]$_63$, and [C I] where the cooling rate depends on the gas density.
Thus, we can constrain the gas density $n_{\mathrm{H}}$ and far-UV incident radiation field $U_{\mathrm{UV}}$ in PDRs from the dust luminosity ($\SI{8}{\mu m}<\lambda<\SI{1000}{\mu m}$) and the ratio of CO to [C I] luminosity.
We perform the PDR modeling for our two samples, M0417-z365 and R0032-z239, which have both CO(4-3) and [C I](1-0) line data using the {\tt PDRTToolbox} \citep{Pound2008, Pound2011, Pound2023}. 
The model of \cite{Kaufman2006} with the gas metallicity of $Z=1$ and simple plane-parallel geometry is assumed.

\begin{figure}[htb]
\begin{center}
    \includegraphics[width=\linewidth]{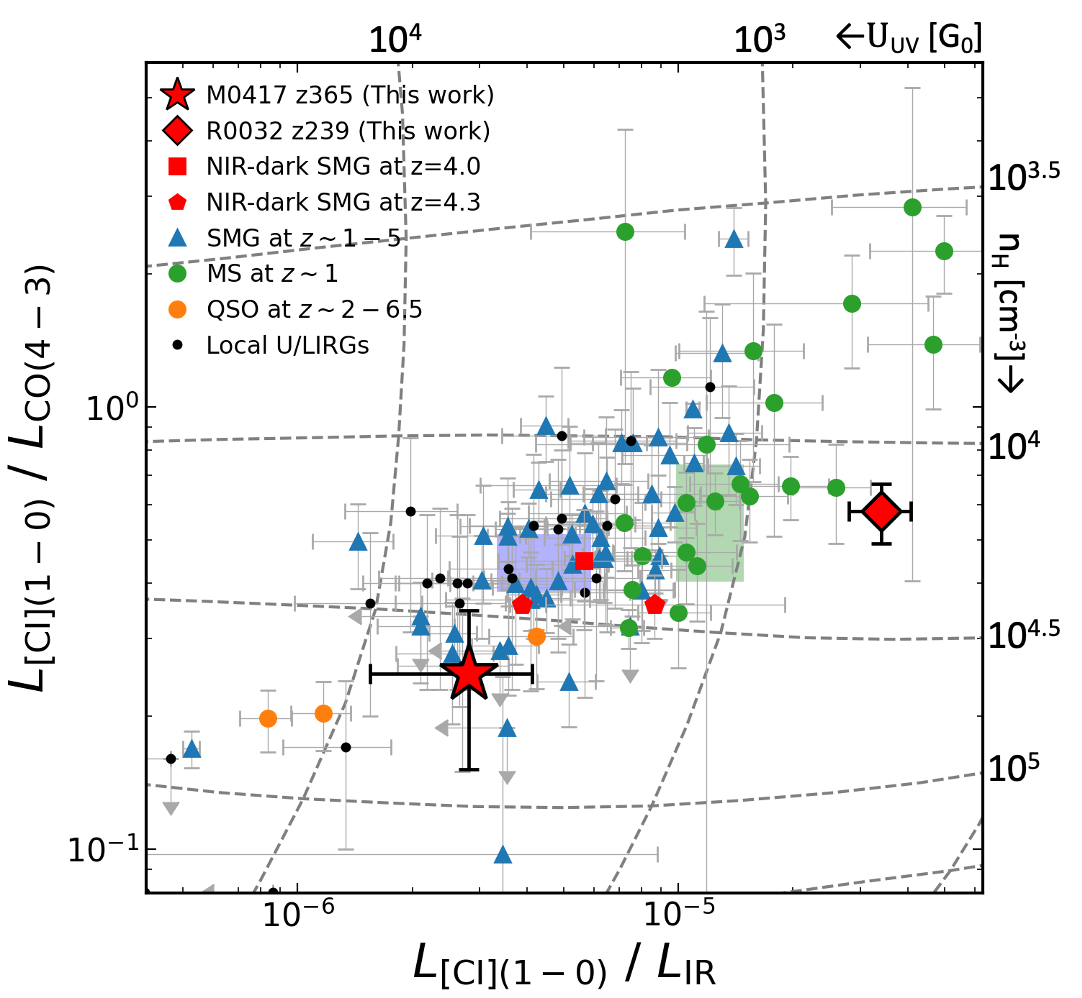}
\end{center}
\caption{$L_\mathrm{[C I](1-0)}/L_\mathrm{CO(4-3)}$ against $L_\mathrm{[C I](1-0)}/L_\mathrm{IR}$ for model and observed values. The horizontal and vertical gray dashed lines represent the model tracks for constant $n_{\mathrm{H}}\,[\si{cm^{-3}}]$ and $U_{\mathrm{UV}}\,[G_0]$, respectively. $G_0=1.6\times 10^{-3}\,\si{ergs.cm^{-2}.s^{-1}}$ is the far-UV radiation field in the vicinity of the Sun \citep{Habing1968}. The red star and diamond are our two samples. The red square and pentagon show NIR-dark SMGs at $z=3.99$ \citep{Umehata2020} and $z=4.26$ \citep{Smail2023}, respectively.  The blue triangle denotes $z\sim1-5$ bright SMGs \citep{Alaghband2013, Bothwell2017, Harrington2021, Birkin2021, Hagimoto2023}. The green and orange circles represent the MS SFGs at $z\sim1$ \citep{Bourne2019, Valentino2020} and $z\sim2-4$ quasar \citep{Alaghband2013, Bothwell2017}. The black dot shows local U/LIRGs \citep{Michiyama2021}. The arrows indicate the $3\sigma$ upper limits. The green and blue shaded region represents the mean line luminosity ratios of MS and SMGs with $2\sigma$ uncertainty derived in \cite{Valentino2020}, respectively.
\label{fig:pdr}}
\end{figure}

Figure \ref{fig:pdr} shows the model predictions for $L_\mathrm{[C I](1-0)}/L_\mathrm{CO(4-3)}$ against $L_\mathrm{[C I](1-0)}/L_\mathrm{IR}$ together with the observed data of our two samples, previously reported NIR-dark SMGs at $z=3.99$ \citep{Umehata2020} and $z=4.26$ \citep{Smail2023}, $z\sim1-5$ bright SMGs \citep{Bothwell2017, Harrington2021, Valentino2020, Hagimoto2023}, MS SFGs at $z\sim1$ \citep{Valentino2020}, $z\sim2-6.5$ QSOs \citep{Alaghband2013,Bothwell2013}, and local U/LIRGs \citep{Michiyama2021}.
While dust continuum and [C I](1–0) emission are generally optically thin, low-$J$ CO emission is optically thick. Thus, we increase the observed CO(4-3) luminosity by a factor of two to incorporate this effect.
M0417-z365 has a lower [C I]/CO and [C I]/FIR luminosity ratio than R0032-z239 and is located closer to other SMGs than MSs or R0032-z239, indicating a higher gas density and stronger radiation field. 
However, we note that the actual ISM would consist of a range of molecular clouds that have different properties like density and radiation field rather than a simple plane-parallel model. Since our measurements are integrated over the whole galaxy, the obtained results should be treated qualitatively.

\section{Discussion}\label{sec:discussion}
In this section, we discuss how our three normal dust-obscured galaxies relate to the broader context of galaxy evolution, primarily based on the properties of molecular gas and the morphology of the star-forming regions.
\subsection{M0417-z365 and R0032-z239: compact dust-obscured galaxies}
Both M0417-z365 and R0032-z239 are main sequence star-forming galaxies and they are characterized by having compact star-forming regions of $\sim 0.6$ kpc, comparable to those of NIR-dark SMGs in literature.
The gas depletion time and molecular gas fraction of M0417-z365 seem to be lower than the scaling relation, although those gas properties are consistent with the scaling relation within the margin of error when $\alpha_{\mathrm{CO}}$ is 3.6.
We first state an interpretation of the former case.
\cite{Elbaz2018} and \cite{Gomez2022} focus on a galaxy population that lies on the MS but has gas depletion times shorter than the scaling relations like starburst (SB) galaxies. 
They term these galaxies "SBs in the MS" and report that they tend to have a smaller gas fraction and higher dust temperature (but see also \citealt{Jin2019, Jin2022, Xiao2023}).
Furthermore, the authors argue that the compactness of the dust continuum area correlates with these characteristics: As the star-forming region becomes more compact compared to the stellar disk, the gas depletion timescale and gas fraction become lower (see also \citealt{Jimenez2019}).
To phenomenologically explain this galaxy population, \cite{Gomez2022} propose a scenario in which the galaxy exhausts its gas from the outside and, through some mechanism, gradually becomes more compact. 
However, this is not confirmed observationally \citep{Ciesla2022}.

Next, we present an interpretation of the latter case.
The molecular gas properties of R0032-z239 are also aligned with the scaling relation when $\alpha_{\mathrm{CO}}$ close to 1.
This case can be interpreted as being in a compaction phase of a steady galaxy evolution scenario for MS galaxies, proposed by cosmological simulations (e.g., \citealt{Zolotov2015, Tacchella2016a}). 
In this scenario, the galaxy experiences compaction of gas and star formation triggered by an intense gas inflow event such as minor mergers or counter-rotating streams. This phase is then followed by inside-out quenching driven by outflows and central gas depletion. 
These processes are repeated until the halo mass grows sufficiently to halt cold gas accretion from the intergalactic medium. 
Throughout these cycles, the galaxy remains MS on the SFR--$M_*$ plane and upholds the scaling relations for gas depletion time and gas fraction.

Both M0417-z365 and R0032-z239 have compact star-forming regions, which aligns with prior studies suggesting a correlation between compactness and the strength of dust attenuation \citep{Smail2021}. 
However, based only on our current results, it is unclear whether this compactness is the actual cause of the strong dust attenuation observed in these galaxies. 

\subsection{R0032-z299: a normal DSFG with edge-on disk}
R0032-z299, the faintest source in our sample, is a LIRG-class MS SFG and its gas depletion time and gas fraction are consistent with the scaling relation, which suggests that this galaxy is under secular evolution.
This source has a dust continuum that is roughly twice as extended as the other two samples and NIR-dark SMGs in the literature. 
Its infrared surface brightness of $\Sigma_{\mathrm{IR}}=2.8\substack{+2.6 \\ -1.3}\times 10^{10}\,L_\odot\si{kpc^{-2}}$ is more than an order of magnitude lower.
The FIR size--luminosity relation at $z\gtrsim2$ is not well constrained at the faint end of $L_{\mathrm{IR}}<10^{12}L_\odot$. 
While \cite{Fujimoto2017} postulates in their statistical ALMA study that fainter galaxies tend to be smaller, \cite{Gonzalez-Lopez2017} and \cite{Rujopakarn2016} report that, despite large scatter, the FIR sizes of fainter DSFGs at $z\sim1-2$ are either comparable to or exceed those of brighter SMGs. In the local universe, the IR-size relation has been investigated down to $L_{\text{IR}} \sim 10^{10} L_\odot$, suggesting that galaxies with lower $L_{\text{IR}}$ tend to have larger sizes \citep{McKinney2023ApJ...955..136M}.

\cite{Gullberg2019} stacks dust continuum of SCUBA-2 detected bright SMGs to find not only a central compact (sub-kpc) component but also a fainter extended component extending about a few kpc associated with (unobscured) star and molecular gas. 
The infrared surface brightness of the compact and extended components are about $(6-10)\times 10^{11}$ and $1\times 10^{10}\,L_\odot\,\mathrm{kpc^{-2}}$, respectively. 
As the SFR increases, the infrared surface brightness of the compact component shows a rising trend, while that of the extended component remains constant. 
Based on this result and simulation works (e.g., \citealt{Hopkins2013}), the authors argue that the two components are physically decoupled. 
They suggest that the compact component likely represents a merger-induced starburst triggered by radial gas inflow that loses its angular momentum. 
Meanwhile, the extended component seems to be a regular disk that had formed before the merger.

The infrared surface brightness of R0032-z299 is comparable to that of the extended component identified by \cite{Gullberg2019} and those of faint ($L_{\mathrm{IR}}\lesssim 10^{12}L_\odot$) DSFGs \citep{Rujopakarn2016,Gonzalez-Lopez2017}.
Together with the normal molecular gas properties, we interpret that R0032-z299 is in a secular evolution phase prior to an outside-in transformation caused by mergers or violent disk instability. 
Another characteristic of this source is its elongated projected axis ratio of 0.17, the most elongated of our three samples.
If we assume axisymmetry, R0032-z299 is likely to be a dusty edge-on disk, although there is another case that even a face-on galaxy can appear elongated if we are viewing a bar-like structure (e.g., \citealt{Hodge2019, Huang2023b}).
Nonetheless, to explain the origin of the strong dust attenuation ($A_V\sim4$), we interpret that this galaxy has an extended dusty edge-on disk, which causes the stellar light to be strongly attenuated by the large dust column density \citep{Wang2018}.
This interpretation suggests that even normal dusty star-forming galaxies can be strongly dust-obscured and missed in optical/NIR observations, depending on the orientation to the observer.
Recently, a {\it JWST} study discovered a {\it HST}-dark galaxy population with similar properties to R0032-z299: massive ($M_*\gtrsim 10^{10}\,M_\odot$), extended ($R_{\mathrm{e,F444W}}\sim1-2$ kpc), and edge-on disk galaxies \citep{Nelson2023}.
We think that these {\it JWST}/NIRCam selected {\it HST}-dark galaxies and R0032-z299 are the same population and that our ALMA observations directly capture the dust responsible for the attenuation.
These were identified in the early observations of {\it JWST}, suggesting that such a galaxy population is not uncommon.
The geometric orientation--whether a galaxy is observed from an edge-on angle or not--would not be the only factor of the strong dust attenuation \citep{Lorenz2023, Gomez-Guijarro2023, Fujimoto2023b}, but in massive galaxies where a disk is present, the edge-on disk orientation may be a significant factor contributing to dust attenuation. 
However, it should be noted that a recent simulation study predicts that edge-on disks tend to have less dust attenuation because stellar light escapes from the edges of the dust distribution \citep{Cochrane2023}.
In the near future, we expect that deep multi-band observation with a high angular resolution by {\it HST}, {\it JWST}, and ALMA will statistically reveal the detailed properties of dust-obscured galaxies and the origins of strong dust attenuation by directly comparing the spatial distribution of stars and dust.
It would also be important to break the degeneracy between dust mass and dust temperature by observing the dust continuum with multiple bands.

\section{Summary}\label{sec:summary}
We present spectroscopic redshifts and physical characterizations of three triply lensed dust-obscured galaxies that are originally detected in the ALCS program. 
The multiple images are apparently bright at \SI{1.2}{mm} band but intrinsically faint ($S_{\SI{1.2}{mm}}<\SI{1}{mJy}$) compared with classical SMGs due to the lens magnifications.

We perform ALMA band 3/4 line scan follow-up observations and successfully confirm the spectroscopic redshifts to be $z=3.6517$, 2.3911, and 2.9853 for M0417-z365, R0032-z239, and R0032-z299, respectively.
We also conduct follow-up observations toward the triply lensed sources in M0417 with the SCUBA-2/450, 850 $\mu$m and FourStar/K-band to collect more multi-band photometries.
Together with archival {\it HST}, {\it Spitzer}/IRAC, {Herschel}/SPIRE, VLT/HAWK-I data, the stellar masses, SFRs, and visual extinctions are derived from panchromatic SED fitting.
Thanks to a total magnification factor of up to $\sim15$, we were able to detect faint CO emission lines and estimate the gas mass.

Our three samples are located in the MS region in the $M_*$--SFR plane, with lower $M_*$ and SFR than classical SMGs.
They show a variety in molecular gas depletion time, gas fraction, and spatial extent of dust.
M0417-z365 and R0032-z239 have compact 1.2 mm dust size of $\sim 0.6$ kpc, comparable to those of previously reported NIR-dark SMGs, following a proposed trend that the more compact the dust continuum size, the higher the $A_V$.
On the other hand, the dust continuum of R0032-z299 is roughly twice as extended as the other two samples and NIR-dark SMGs in the literature. Its infrared surface brightness is more than an order of magnitude lower than the other two samples and SMGs, which suggests that R0032-z299 is a different population.
The molecular gas depletion time and gas fraction of R0032-z299 are consistent with a scaling relation for star-forming galaxies, suggesting this galaxy undergoes a secular evolution phase.
R0032-z299 has an elongated projected axis ratio of $\sim 0.17$, the most elongated of our three samples.
This suggests that the edge-on dusty disk configuration causes strong dust attenuation.
If so, even normal DSFGs can be strongly dust-obscured and missed in optical/NIR observations, depending on the orientation to the observer.

\section*{Acknowledgments}
We thank Masato Hagimoto for providing us with the [C I](1-0)/CO(4-3)/FIR data. 
We appreciate Takeo Minezaki for constructive comments and suggestions.
We also sincerely thank the anonymous referee for constructive comments and suggestions, which have improved the quality of this paper.
This paper makes use of the following ALMA data: ADS/JAO.ALMA\#2018.1.00035.L, \#2019.1.00949.S, and \#2021.1.01246.S. ALMA is a partnership of ESO (representing its member states), NSF (USA) and NINS (Japan), together with NRC (Canada), NSTC and ASIAA (Taiwan), and KASI (Republic of Korea), in cooperation with the Republic of Chile. The Joint ALMA Observatory is operated by ESO, AUI/NRAO and NAOJ. 
This work is based (in part) on data collected at the Magellan Telescopes, with the support from the ALTA program of ASIAA.
Data analysis was in part carried out on the Multi-wavelength Data Analysis System operated by the Astronomy Data Center (ADC), NAOJ.
This research was supported by FoPM, WINGS Program, the University of Tokyo.
AT was supported by the ALMA Japan Research Grant of NAOJ ALMA Project, NAOJ-ALMA-343.
AT also acknowledges the support by JSPS KAKENHI Grant Number 24KJ0562.
KK acknowledges the support by JSPS KAKENHI Grant Number JP17H06130 and JP22H04939, and the NAOJ ALMA Scientific Research Grant Number 2017-06B. 
HU appreciates the support by JSPS KAKENHI Grant Number 20H01953, 22KK0231, and 23K20240.
WHW and ZKG acknowledge support by NSTC grant 111-2112-M-001-052-MY3.
FS acknowledges JWST/NIRCam contract to the University of Arizona NAS5-02015.
RU appreciates the support by JSPS KAKENHI Grant Number 22J22795.
FEB acknowledges support from ANID-Chile BASAL CATA FB210003, FONDECYT Regular 1200495,
and Millennium Science Initiative Program  – ICN12\_009.

%

\vspace{5mm}
\facilities{HST (ACS and WFC3), Magellan (FourStar), VLT (HAWK-I), Spitzer (IRAC), Herschel (PACS), JCMT (SCUBA-2), ALMA}


\software{Astropy \citep{astropy:2022},
          Glafic \citep{Oguri2010},
          CASA \citep{McMullin2007},
          Photutils \citep{photutils},
          SExtractor \citep{Bertin1996},
          PSFEx \citep{Bertin2011},
          CIGALE \citep{Boquien2019},
          PDRToolbox \citep{Pound2023},
          Dynesty \citep{Speagle2020},
          }


\appendix

\section{SED fitting}
\begin{figure*}[ht!]
\centering
\includegraphics[width=\linewidth]{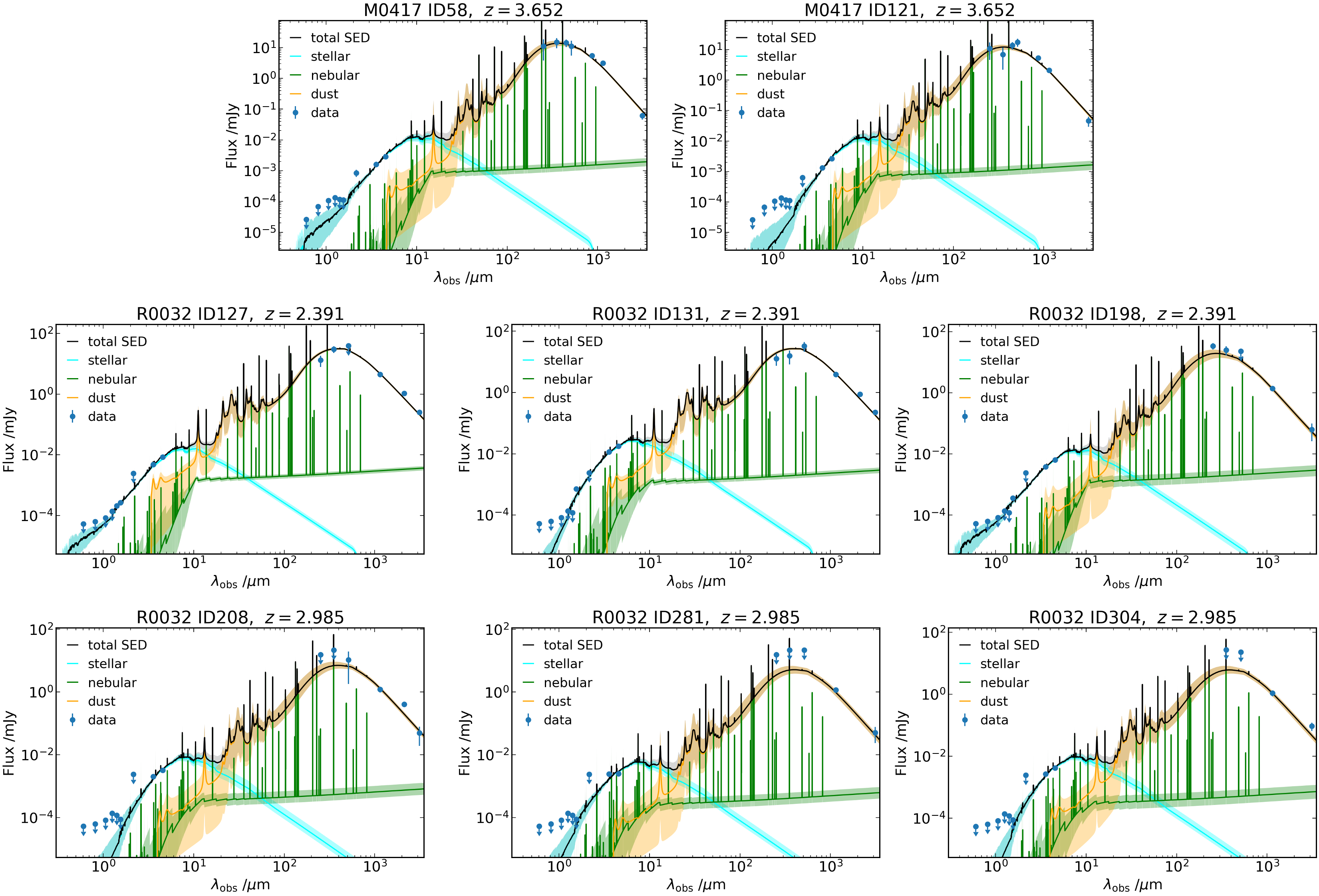}
\caption{Median (solid lines) and 16th–84th percentile range (shaded regions) SEDs of our samples. The arrows indicate the $3\sigma$ flux upper limits used in the SED fitting. 
\label{fig:sed}}
\end{figure*}

Table \ref{tab:param_cigale} lists the parameter ranges used for our sed fitting. 
Figure \ref{fig:sed} shows the sed fit results for all our sources.

\begin{table*}[tbh]
\caption{Parameter ranges used for the SED fitting with {\tt CIGALE}.\label{tab:param_cigale}}
\begin{tabularx}{\linewidth}{@{}lll}
 \hline\hline
 Parameter & Symbol &Value\\
 \hline
\multicolumn{3}{c}{SFH ({\tt nonparametric}; \citealt{Leja2019})} \\ 
{Number of constant SFH bins}&    &   6   \\
{Prior}&    &  Continuity  \\
&    &    (SFR$_n$/SFR$_{n+1}\sim$ Student's-t distribution)   \\

\hline 
\multicolumn{3}{c}{SSP ({\tt bc03}; \citealt{Bruzual2003})} \\ 
{Initial mass function} & IMF &  \cite{Chabrier2003}   \\
{Metallicity of the stellar model} & $Z_{\text{star}}$ & 0.02  \\
\hline 
\multicolumn{3}{c}{Nebular emission ({\tt nebular})} \\ 
{Ionisation parameter} & $\log U$ &  -2.0   \\
{Gas Metallicity} & $Z_{\text{gas}}$ & 0.02  \\
{Electron density} & $n_e$ [\si{cm^{-3}}] & 100 \\
\hline 
\multicolumn{3}{c}{Dust attenuation ({\tt dustatt\_calzleit}; \citealt{Calzetti2000})} \\ 
{Color excess of the stellar continuum light for the young population} & E(B-V)$_*$ & 0--10 \\
{Reduction factor to apply E(B-V)$_*$ for the old population} &  & 0--1 \\
{UV bump amplitude} & & 0 \\
{Slope of the power law modifying the attenuation curve}& $\delta$ & -0.7--0.4   \\ 
\hline 
\multicolumn{3}{c}{Dust emission ({\tt dl14}; \citealt{Draine2014})} \\ 
{Mass fraction of Polycyclic Aromatic Hydrocarbons (PAH)}& $q_{PAH}$ & 0.47, 1.12, 2.50, 3.19, 3.90   \\
{Minimum radiation field}& $U_{\text{min}}$ & All possible values above 5.0   \\
{Power-law index of starlight intensity distribution}& $\alpha$ & All possible values (1.0--3.0)   \\
{Fraction illuminated from $U_{\text{min}}$ to $U_{\text{max}}$} & $1-\gamma$ & 0--1 \\
\hline
\end{tabularx}
\tablecomments{We use flat prior distributions for all parameters. \( t_z \) represents the age of the universe at redshift \( z \).}
\end{table*}

\section{Intrinsic dust continuum size} \label{appe:resolution}
In section \ref{sec:morphology}, we reconstructed the intrinsic source morphology of the 1.2 mm continuum from ALCS observations using a MCMC approach. Here, we discuss an alternative method to demonstrate that the ALCS observations, with a resolution of $\sim1\arcsec$, are capable of spatially resolving structures with $r_e\sim0.5$ kpc. 

In principle, it is possible to resolve sources smaller than the beam if the beam profile is well-defined, the SNR is sufficiently high, and some assumptions about the source light profile are made. \cite{Simpson2015} demonstrated that when the source size is close to the synthesized beam size and a peak SNR is more than $\sim 10$, source sizes can be recovered with a precision of $\sim30$\%. \cite{Tsujita2022PASJ...74.1429T} demonstrated that, in the presence of strong lensing, it is possible to measure source sizes even when they are smaller than the effective beam size.
The minimum radius $r_{\text{min}}$ that can be reliably measured can be expressed as follows \citep{MV2012A&A...541A.135M, Gomez2022}: $r_{\text{min}}\simeq0.44~ \theta_{\text{beam}}/\sqrt{\text{SNR}}$, where $\theta_{\text{beam}}$ is the FWHM of the beam. Note that the effective angular resolution is improved by a factor of $\sqrt{\mu}$ due to the strong lensing. For example, M0417-z365.ID46 has a magnification factor of $\mu = 5.1$ (Table \ref{tab:line_property}) and a peak SNR of $\sim50$ \citep{Fujimoto2024ApJS..275...36F}. Substituting these values into the above equation yields $r_{\text{min}}\sim0.03\arcsec$, which corresponds to $\sim0.2$ kpc at $z=3.652$. This calculation demonstrates that this source is indeed resolvable with our observations. 
Similarly, for M0417-z365.ID58 and M0417-z365.ID121, we obtain $r_{\text{min}} \sim 0.3$ kpc and $\sim 0.4$ kpc at $z = 3.652$, respectively. For R0032-z239.ID127, R0032-z239.ID131, and R0032-z239.ID198, the calculations yield $r_{\text{min}} \sim 0.2$ kpc, $\sim 0.2$ kpc, and $\sim 0.5$ kpc at $z = 2.391$, respectively. Likewise, for R0032-z239.ID208, R0032-z239.ID281, and R0032-z239.ID304, we obtain $r_{\text{min}} \sim 0.4$ kpc, $\sim 0.5$ kpc, and $\sim 0.6$ kpc at $z = 2.985$. These results confirm that all sources in our sample are spatially resolved.


\bibliographystyle{aasjournal}
\bibliography{main}



\end{document}